%% file: main.tex
\documentclass[journal]{IEEEtran}
\usepackage[english = american]{csquotes} 
\usepackage{url,multirow,comment,hyperref,amsmath,graphicx,amsmath,subfigure,listings,tabularx,psfrag,wrapfig,cite,verbatim,array, algorithm, algorithmicx, algpseudocode}
\usepackage[
locale = DE 
]{siunitx} 
\usepackage{caption} 
\usepackage{xcolor}
\begin{document}


\title{Graph-Based Intrusion Detection System for Controller Area Networks}

\author{
Riadul~Islam,~\IEEEmembership{Member,~IEEE},
Rafi Ud Daula Refat,~\IEEEmembership{Student Member,~IEEE},
Sai Manikanta Yerram,
~and  Hafiz~Malik,~\IEEEmembership{Senior Member,~IEEE}

\thanks{R Islam and S M Yerram are with the Department 
of Computer Science and Electrical Engineering, University of Maryland, Baltimore County, Baltimore, 21250 USA e-mail: {\{riaduli, ty16571\}@umbc.edu}.}
\thanks{R Refat and H Malik are with the Department 
	of Electrical and Computer Engineering, University of Michigan, 
	MI, 48128 USA e-mail: {\{rerafi, hafiz\}@umich.edu}}
\thanks{This work was supported in part by the UMBC startup grant and by the Ministry of Education in Saudi Arabia under the grant DRI-KSU-934.}
\thanks{Copyright (c) 2020 IEEE. Personal use of this material is permitted. 
However, permission to use this material for any other purposes must be 
obtained from the IEEE by sending an email to pubs-permissions@ieee.org.}
}
\markboth{IEEE Transactions on Intelligent Transportation Systems}
{Shell \MakeLowercase{\textit{et al.}}: ??????}

\newcommand{\fixme}[1]{{\Large FIXME:} {\bf #1}}

\maketitle


\begin{abstract}
The controller area network (CAN) is the most widely used intra-vehicular communication network in the automotive industry. 
Because of its simplicity in design, it lacks most of the requirements needed for a security-proven 
communication protocol. However, a  safe and secured environment is imperative for autonomous as well as connected vehicles. 
Therefore CAN security is considered one of the important topics in the automotive research community. In this paper, we propose a four-stage intrusion detection system that uses the chi-squared method and can detect any kind of strong and weak cyber attacks in a CAN. This work is the first-ever graph-based defense system proposed for the CAN. Our experimental results show that we have a very low 5.26\% misclassification for denial of service (DoS) attack, 10\% misclassification for fuzzy attack, 4.76\% misclassification for replay attack, and 
no misclassification for spoofing attack. In addition, the proposed methodology exhibits up to 13.73\% better accuracy compared to 
existing ID sequence-based methods. 
\end{abstract}
\IEEEpeerreviewmaketitle

\input{intro}

\input{background}

\input{proposed_methodology}



\input{analysis}


\input{conclusion}



\bibliographystyle{IEEEtran}
\bibliography{main}

\vspace{-2.0cm}
\begin{IEEEbiography}[{\includegraphics[width=1in,height=1.25in,clip,keepaspectratio]{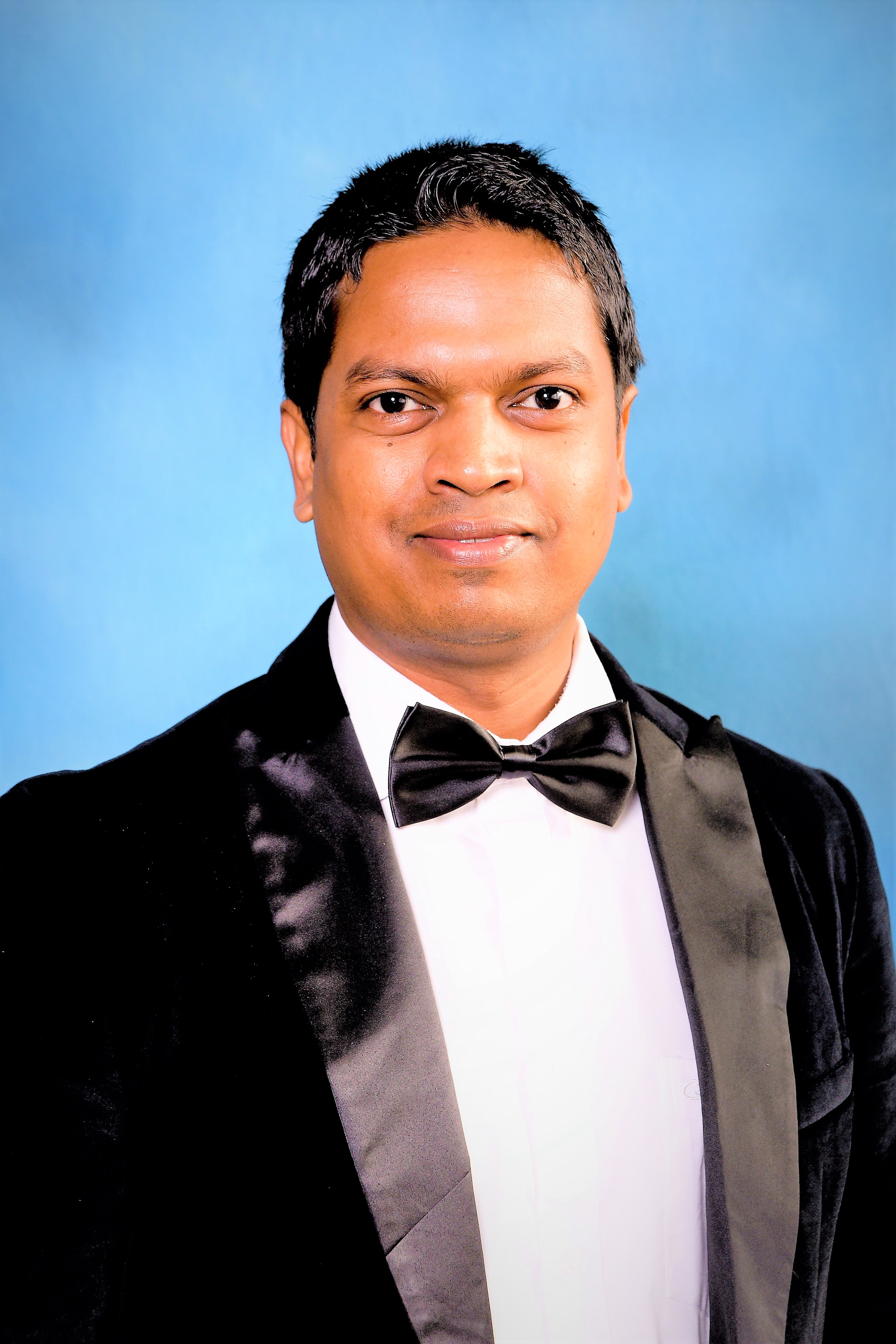}}]{Riadul Islam}
	is currently an assistant professor in the
Department of Computer Science and Electrical Engineering at the University of Maryland, Baltimore County. 
In his Ph.D. dissertation work at UCSC, Riadul
designed the first current-pulsed flip-flop/register that resulted in the 
first-ever one-to-many current-mode clock distribution networks for
high-performance microprocessors. From 2017 to 2019, he was an Assistant
Professor with the University of Michigan, Dearborn MI, USA. From 2007 to 2009, he worked as a
full-time faculty member in the Department of Electrical and Electronic
Engineering of the University of Asia Pacific, Dhaka, Bangladesh. He is a
member of the ACM, IEEE, IEEE Circuits and Systems (CAS) society, and IEEE Solid-State Circuits (SSC) Society. 
He is a member of the Cybersecurity Center for Research, Education, and Outreach at the UM-Dearborn. He holds two US 
patent and several IEEE/ACM/Springer Nature journal and conference publications 
in TVLSI, TCAS, JETTA, DAC, ISCAS, MWSCAS, ISQED, and ASICON.  His  current  research
interests include  digital, analog, and mixed-signal CMOS ICs/SOCs for a 
variety of applications; verification and testing techniques for analog, 
digital and mixed-signal ICs; hardware security; CAN network; CAD tools for design and analysis of
microprocessors and FPGAs; automobile electronics; and biochips. 
He is an editorial board member of Semiconductor Science and Information Devices journal.
\end{IEEEbiography}
\vspace{-0.0cm}

\vspace{-1.0cm}
\begin{IEEEbiography}[{\includegraphics[width=1in,height=1.25in,clip,keepaspectratio]
		{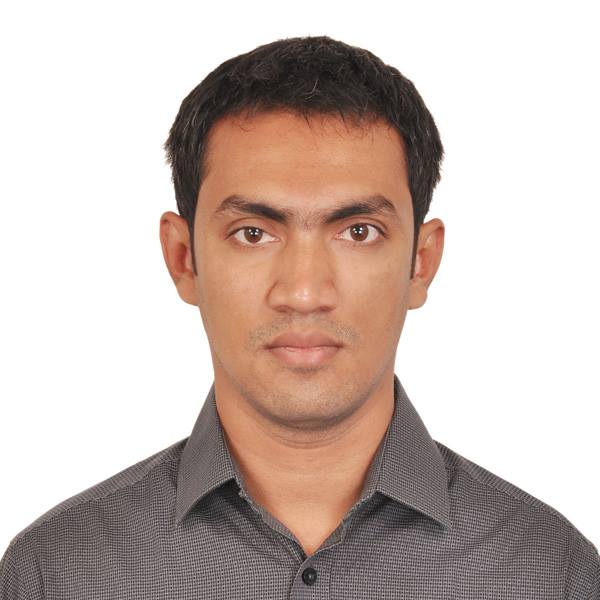}}]{Rafi Ud Daula Refat}
received the BSc degee in Computer Science \& Engineering from Rajshahi University of Engineering \& Technology. 
He is currently pursuing the Ph. D. degree with the department of Electrical \& Computer Engineering, University of Michigan - Dearborn, MI, USA.
His research interest focus on cybersecurity and data analysis. 
\end{IEEEbiography}
\vspace{-2.0cm}

\vspace{1.0cm}
\begin{IEEEbiography}[{\includegraphics[width=1in,height=1.25in,clip,keepaspectratio]
		{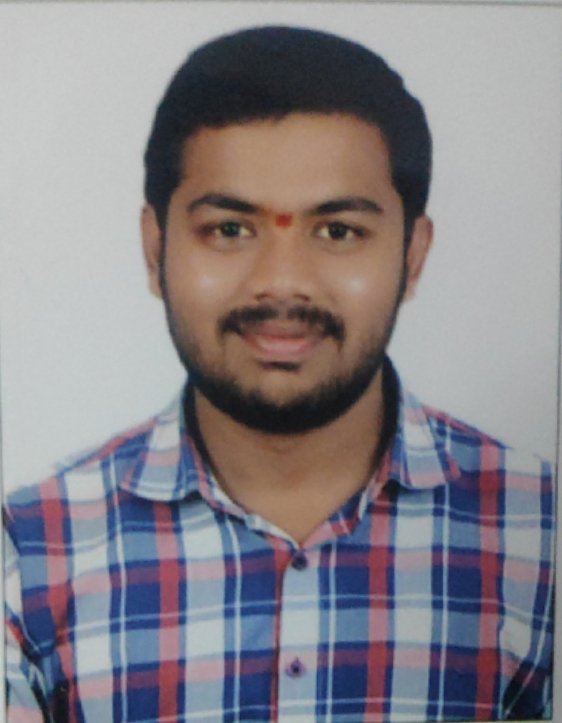}}]{Sai Manikanta Yerram}
received bachelor’s degree in Information Technology from Sreenidhi Institute of Science and Technology, Hyderabad, India in 2019. He is a Grad student pursuing Master of Professional Studies in Data Science at the University of Maryland, Baltimore County. His current areas of interest are focused on  Machine Learning and their applications. 
\end{IEEEbiography}
\vspace{4.5cm}

\vspace{-1.5cm}
\begin{IEEEbiography}[{\includegraphics[width=1in,height=1.25in,clip,keepaspectratio]{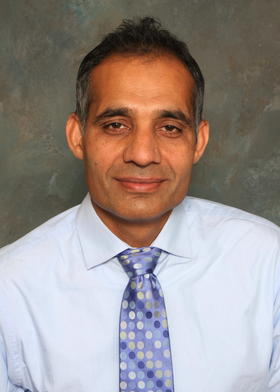}}]{Hafiz Malik}
is an Associate Processor in the Electrical and Computer Engineering (ECE) Department at University of Michigan, Dearborn. His research in the areas of automotive cybersecurity, IoT security, sensor security, multimedia forensics, steganography/steganalysis, information hiding, pattern recognition, and information fusion is funded by the National Science Foundation, National Academies, Ford Motor Company, and other agencies. He has published more than 80 papers in leading journals, conferences, and workshops. He is a founding member of the Cybersecurity Center for Research, Education, and Outreach at the UM-Dearborn. He is also serving a member of Scientific and Industrial Advisory Board (SIAB) of the National Center of Cyber Security Pakistan. He is a member, MCity Working Group on Cybersecurity, since 2015. He is a member of Working Group for IEEE Project 1912--Standard for Privacy and Security Architecture for Consumer Wireless Devices. He is also on the Review Board Committee of IEEE Technical Committee on Multimedia Communications (MMTC). 
\end{IEEEbiography}
\vspace{-0.5cm}

\end{document}

%% file: intro.tex
\section{Introduction}
\label{sec:intro}

Autonomous or self-driving vehicles are cars or trucks for which human interaction is not needed in driving. 
Also known as driverless vehicles, they are equipped with various types of sensors, actuators, a high-performance 
computing system, and software. Although a fully autonomous car is still not a reality, the demand for 
partially autonomous cars with various levels of self-automation is very high. 
It seems that the successful development of a fully autonomous vehicle will change the world's overall transportation 
system and economy. Hypothetically, it will provide more safety and reduce the accident rate, as every year 
many accidents occur only because of mistakes made by the human driver. In addition, an autonomous vehicle 
can be considered a blessing for the disabled person. To turn the dream into a reality, all the original 
equipment manufacturers (OEMs) are working on this technology, and it is hoped that we will experience 
the highest level of autonomous features within the next several years.

However, the key consideration for autonomous vehicles should be providing protection against 
cyber attackers. As the autonomous vehicle will totally depend on the software, sensors, and 
third-party signals to operate, one can expect that it will catch the attention of hackers. 
The impact of cyber or physical attacks performed by intruders can include the disclosure of
vehicle or driver information like location, gender, number of passengers currently riding, etc.
In order to provide safety, 
it is important to establish a strong protection mechanism not only in vehicle-to-vehicle 
communication but also in intra-vehicle communication. For an autonomous vehicle, both the 
inter-vehicle and intra-vehicle communications are controlled by electronic control units (ECUs). 
ECUs are called the brain of the self-driving car and are responsible for taking 
real-time decisions, so it is important that they exchange information among themselves over a secured communication channel. 

For the intra-vehicle communication channel, controller area network (CAN) technology 
is considered the {\it de facto} standard among the car embedded systems~\cite{Bosch:1991}. 
However, the CAN protocol has some serious security breaches in its core. The protocol actually 
works similarly to a broadcasting system, in that contains no mechanism for checking the 
identity of the sender. Several researchers have tried to provide solutions to increase 
the security of the CAN bus~\cite{Cho:2016,Lee:2017,Woo:2015,Tobias:2008}. Most of them work for a certain type of 
attack situation. Currently, the increasing amount of research work on autonomous 
vehicles has inspired us to work on detecting anomalies in the CAN bus for autonomous vehicles.

Several researchers have proposed different solutions for defending against cyber attacks 
in a vehicular system. An attack can be performed through either a weak or a strong agent~\cite{Ji:2018}. 
A weak agent can spoof the CAN bus by injecting messages with high priority 
(ID 0000, denial of service (DoS) attack~\cite{Lin:2012}) or with any arbitration ID (spoofing~\cite{Iehira:2018} or 
Fuzzy attack~\cite{Islam:2020,Lee:2017}). 
On the other hand, a strong agent uses two attackers at the same time to perform attacks. In that case, 
one attacker tries to suspend the targeted ECU's communication and the other attacker sends 
a CAN bus message with the targeted ECU's arbitration ID~\cite{Lee:2017,Lin:2012}. 
Unlike conventional methodologies, we try to 
find complex relationships about CAN bus data using graph theory.

Graph-based anomaly detection techniques are used widely in 
industries like finance, fraud detection, computer and social networking, data center 
monitoring, etc.~\cite{Akoglu:2015,Nong:2001}. 
The unusual substructure of a graph can be used as a 
flag or sign of an anomaly. First, we construct a set of graphs from CAN bus data and then 
search for unusual behavior to flag as an anomaly. Our experimental results 
showed significant success in detecting anomalies with this approach.

In particular, the major contributions in this work are:
\begin{itemize} \renewcommand{\labelitemi}{$\bullet$}
	\item It is the first-ever graph-based cyber attack defense system for CAN communication.
	\item It is the first chi-squared distribution implementation for detecting attacks in CAN communication.
	\item We propose a four-stage intrusion detection system (IDS) for cyber or physical 
attacks on the CAN bus. Here, we use a graph-based approach to find out patterns in 
the dataset, and the median test and chi-squared test are used to distinguish two data distributions.
	\item Our proposed algorithm can detect attacks without any change in the CAN protocol. 
Therefore, it is applicable to any communication system that uses the CAN protocol.
\end{itemize}

The rest of the paper is organized as follows: Section~\ref{sec:back} discusses the existing CAN; 
statistical hypothesis testing, especially the chi-squared test; and graph properties. At the end of 
this section, we present the related work on anomaly detection in the CAN bus as well as graph-based 
anomaly detection techniques. Section~\ref{sec:proposed_method} describes our proposed solution. 
Section~\ref{sec:analysis} and Section~\ref{sec:conclusion} present details about the experimental results and conclusion, respectively.

%% file: background.tex
\section{Background and Related Works}
\label{sec:back}
The broadcasting nature of a CAN communication system confirms that all the messages that 
are transmitted in the network are accessible by all the connected ECUs. After receiving a 
CAN message, each ECU translates the bit sequence and extracts the necessary components 
like arbitration ID, data, cyclic redundancy check (CRC), etc., and finally the ECU 
decides whether to receive the CAN message or not based on the arbitration ID. 
Apart from indicating the sender of a particular CAN message, the arbitration ID of a 
CAN message is used as a priority during a collision between two or more 
CAN messages. For conflict resolution between two CAN messages, carrier-sense multiple 
access with collision avoidance is used in the CAN bus protocol~\cite{Lee:2017}.  
Although the arbitration ID is used to define the priority and the source to resolve 
conflict, it is incapable of authenticating the origin of that CAN message. This security 
flaw can be used by the inside attacker (who already has access to the CAN bus) and the 
outside attacker (who can gain access using the cellular communication network, etc.) to corrupt the CAN system.

\begin{figure*}[t!]
	\centering
	\vspace{-0.0cm}
	\includegraphics[width = 0.85\textwidth]{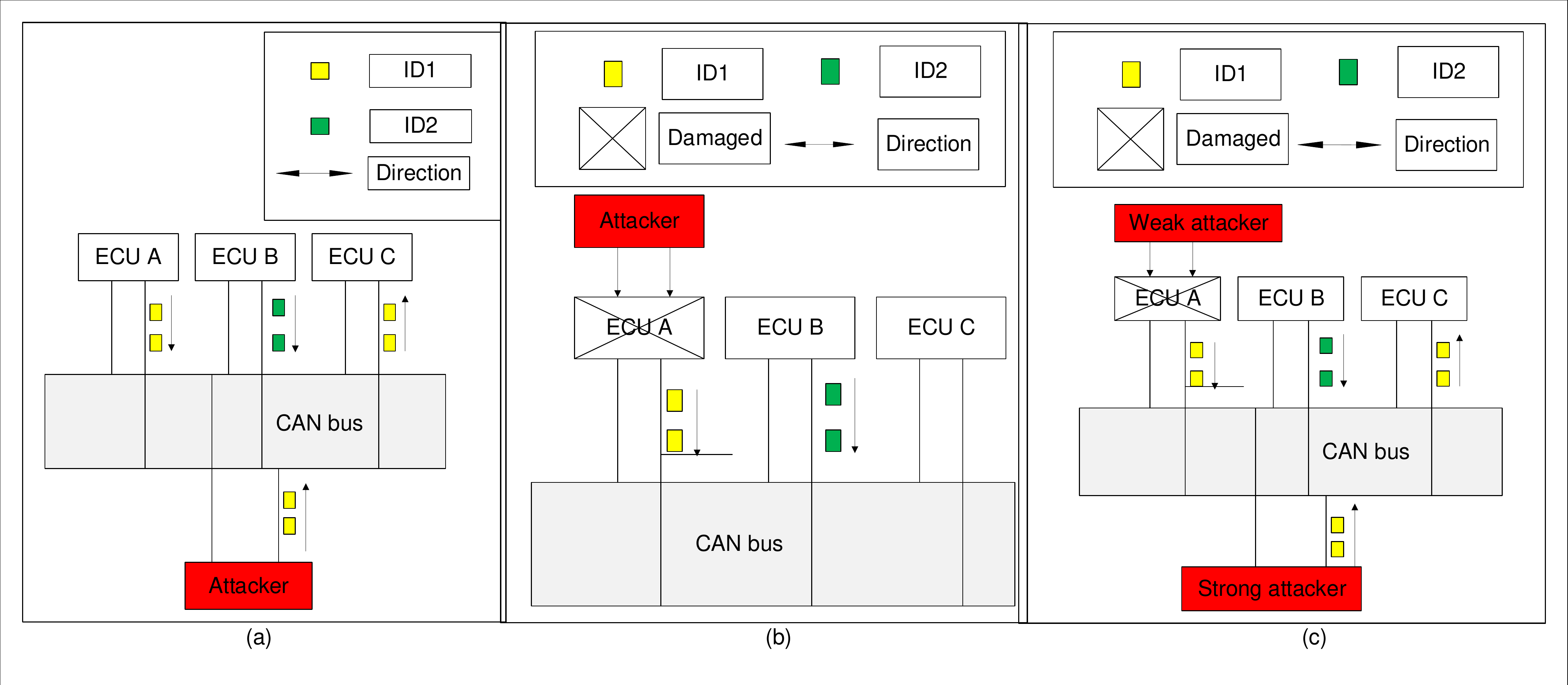}
	\caption {CAN monitoring-based attack mechanisms: (a) in a fabrication attack, the 
		attacker uses a compromised ECU to inject a message with a forged ID; (b) in a 
		suspension attack, the intruder suspends the communication of a compromised ECU; 
		and (c) in a masquerade attack, the intruder uses a weak attacker to suspend one 
		ECU's data communication and uses a strong attacker to send messages mimicking the suspended ECU's ID and frequency.}
	\label{fig:all_attack}
	\vspace{-0.4cm}
\end{figure*}

These attackers can initiate different kinds of attacks on the CAN bus, and we can generalize them as follows:
\begin{itemize} \renewcommand{\labelitemi}{$\bullet$}
	\item Fabrication attack: Through an in-vehicle ECU compromised by a strong attacker, 
	the adversary fabricates and injects messages with forged ID, data length code, and data. 
	Figure~\ref{fig:all_attack}(a) shows an example of the fabrication attack. The objective 
	of this attack is to override any periodic messages sent by a legitimate safety-critical 
	ECU so that their receiver ECUs get distracted or become inoperable. DoS~\cite{Lin:2012}, 
	spoofing~\cite{Iehira:2018}, and fuzzy attacks~\cite{Lee:2017} are some examples of fabrication attacks~\cite{Lin:2012}.
	\item Suspension attack: To mount a suspension attack, the adversary needs only one weakly 
	compromised ECU, and that becomes a weak attacker. The objective of this attack is to 
	stop/suspend the weakly compromised ECUs message transmission, thus preventing the 
	delivery/propagation to other ECUs of the information it acquired~\cite{Tomlinson:2018}. 
	Figure~\ref{fig:all_attack}(b) shows an example of the suspension attack, where a weak attacker 
	suspends ECU A's operation. As a result, this attack affects the performance of 
	various ECUs that utilize certain information from other ECUs to function properly. 
	Therefore, the suspension attack can harm not only the (weakly) compromised ECU itself 
	but also other receiver ECUs. 
	\item Masquerade attack: To mount a masquerade attack~\cite{Lin:2012}, the adversary 
	needs to compromise two ECUs, one as a strong attacker and the other as a weak attacker, 
	as shown in Figure~\ref{fig:all_attack}(c). The adversary monitors and learns which 
	messages are sent at what frequency by the weakly attacked ECU, and then the strong attacker 
	transmits the message with the ID of the compromised ECU at the same frequency. 
	Examples of this attack are replay or impersonation attacks~\cite{Choi:2018}.Moore
\end{itemize}

\subsection{Chi-Squared Test}
\label{subsec:chi}

In our proposed methodology, we use the chi-squared test to detect the anomalous 
CAN data. The chi-squared independence test~\cite{Ugoni:1995}  
is a statistical procedure to test if two categorical distributions belong to the same 
populations or not. It uses the frequency of each category as a factor for distinguishing 
two distributions. In other words, it is called the chi-square goodness of fit because 
in this test, the expected frequencies of all the features of one data distribution are 
extracted and then the findings are compared with the observed frequencies of all 
the features of the second distribution. If the two distributions are significantly 
different, we will call it a reject hypothesis; otherwise, we will call it a null 
hypothesis (if the distributions are the same). The chi-squared test can be described by the following equation,
\begin{equation}
\label{eq:chi1}
X_{DoF}^2 = \sum_{i=0}^{DoF} \frac {(O_i - E_i)^2} {E_i}
\end{equation}
where $DoF$ is the degree of freedom, $X_{DoF}^2$ is the value we will compare against a 
threshold, $O$ is the observed frequency, and $E$ is the expected frequency. The degree of 
freedom in the chi-squared test actually specifies the shape of the distribution. 
It is a numerical value and can be represented by following equation,
\begin{equation}
\label{eq:chi2}
DoF = (i-1) \times (j-1)
\end{equation}
where i is the number of categories and j is the number of rows used in that test. 
The degree of freedom is important for finding out the threshold value from the chi-square 
table of significance. In general, the chi-squared test works well in a categorical larger 
data set, which is our primary motivation for using this test for comparing 
CAN bus-based data distribution. 

The chi-squared test is considered one of the most reliable statistical 
tests now available. Researcher have been relying on the chi-square 
test for more than a hundred years.
Previously, the chi-square test was applied in data correlation~\cite{Steiger:1985}, 
experimental and theory-based probability relationship~\cite{Tallarida:1987}, 
and chiropractic and osteopathic data analysis~\cite{Ugoni:1995}.
In recent years, the chi-squared test has been applied in text 
classification~\cite{Phayung:2011}, and in an IDS with a 
multi-class support vector machine (SVM)~\cite{Thaseen:2017}.
It is a trustable method and has been proven in sectors like medical~\cite{Kobayashi:2013}, 
anomaly-detection~\cite{Nong:2001}, and other kinds of data-analysis problems~\cite{Phayung:2011}. 
Other researchers have used the chi-squared test for real-time detection of navigation system soft failures~\cite{Brumback:1987}.
However, to the best of our knowledge, no previous work used the chi-squared test for anomaly detection in a vehicular network.

\subsection{Graph Properties}
\label{subsec:graph_pro}

A graph is a non-linear data structure consisting of vertices and edges. 
It actually allows representation of the relationship between two vertices. We can 
easily understand the relationship between two vertices from a graph. Over the years, 
researchers extensively used graph properties to solve various problems in 
computer science, operating systems, Google maps, social media, etc. 
Table~\ref{tab:graph_pro} presents graph properties and their significance.

\begin{table} [t!] \large 
	\renewcommand{\arraystretch}{1.5}
	\caption{The proposed methodology uses common graph properties; for example, an edge represents a sequence of CAN messages, while degree represents the number of arbitration IDs sequential with the current ID.}
	\label{tab:graph_pro}
	\centering
	\scalebox{0.65}{
		{\begin{tabular}{|c|c|c|c|c|}
				\hline
				Properties               & Significance                      & In Can bus data      \\
				\hline
				Vertex        & Node of the graph      & Arbitration ID \\
				\hline
				Edge      & Link between nodes     & Arbitration ID sequence    \\
				\hline
				\multirow{2}{*}{Degree} & \multirow{2}{*}{How many neighbours}  & \multirow{2}{*}{How many arbitration IDs}    \\
				                       &              &    are sequential with current ID        \\
				\hline
				\multirow{2}{*}{Cycle}       & \multirow{2}{*}{Loop in the graph}     & \multirow{2}{*}{Loop between the }  \\
				&              &    sequential arbitration IDs       \\
				\hline
				Root      & Starting of the graph           &  The first CAN message   \\
				\hline
				
	\end{tabular}}}
	\vspace{-0.30cm}
\end{table}

\subsection{Related Works}
\label{subsec:related_work}
Due to the widespread use of CAN, there has been a significant amount of work on 
CAN security~\cite{Lin:2012,Iehira:2018,Cho:2016,Lee:2017,Woo:2015,Tobias:2008,Ling:2012,Ji:2018,Fu:2016,Taylor:2015,Müter:2011,Marchetti:2016,Song:2016,Marchetti:2017,Tomlinson:2018,Islam:2020}. 
Researchers incorporate conventional validation techniques 
to identify invalid CAN ID~\cite{Ling:2012}. To identify anomaly, researchers 
build a regular model either considering internal CAN messages or vehicle 
specifications. However, this method can be easily evaded by using existing 
fabrication attacks. Other researchers proposed a decision tree-based detection 
system considering eight physical and cyber features~\cite{Ji:2018}. This method 
uses small-scale robotics vehicle to validate the proposed methodology; 
however, has high detection latency. Another exciting work introduced an 
FPGA-based IDS for vehicular systems. Yet, this 
method is specified for the FPGA platform only, which limits its 
application to other CPU and GPU-based systems~\cite{Fu:2016}.

Other researchers use the periodicity of CAN's message to detect 
anomaly~\cite{Cho:2016,Song:2016}. Empirically, ECUs 
generates CAN message at a specific frequency. As a result, it 
is possible to detect anomalies in inter-arrival time when an 
external attacker injects messages. Another frequency-based 
IDS uses a one-class support vector machine 
to detect anomalies with high accuracy~\cite{Taylor:2015}. However, 
real CAN message prone to variation, and often exhibits inconsistent 
inter-arrival time, reduces the reliability of these schemes.

Some researchers used information theory and proposed an entropy-based 
IDS to detect CAN attacks. The basic idea of these 
systems is normal attack-free CAN messages will have standard or 
stable entropy. On the other hand, the attacked CAN messages will 
have unstable behavior~\cite{Müter:2011,Marchetti:2016}. Another 
exciting approach and closest to our proposed method is arbitration 
ID sequencing-based IDS~\cite{Marchetti:2017}. This method builds a 
transition matrix using a standard CAN dataset and compares attacked 
CAN message ID sequence to detect an attack. This method can detect 
simple impersonate-type attacks; however, it could not detect replay attacks.

%% file: proposed_methodology.tex
\section{Proposed Methodology}
\label{sec:proposed_method}
In this paper, we propose an IDS to secure 
the CAN bus communication system. The proposed methodology uses statistical 
analysis as a basis for detecting anomalies and is divided into several steps. 
The steps are (i) transferring a CAN bus message to a more meaningful graph 
structure, (ii) extracting graph-based features to import for anomaly 
characterization, (iii) constructing a hypothesis based on the safe population 
window, and (iv) comparing the test population window to the base population. 
This section is organized as follows: Section~\ref{subsec:proposed_ids} 
contains the proposed algorithms and Section~\ref{subsubsec:graph_features} 
contains the exploratory data analysis. 

\subsection{Proposed Intrusion Detection Methodology} 
\label{subsec:proposed_ids}
In this section, we will first define two terms used in our proposed algorithm and 
then detail our proposed methodology. The terms are:
\begin{itemize} \renewcommand{\labelitemi}{$\bullet$}
	\item {\bf Window:} A range of raw CAN bus messages will be called a single 
	window. In our proposed methodology, we consider 200 messages to be window size. 
	In the results section, we will discuss it in detail.
	\item {\bf Population Window:} A population window is a set of windows. 
	It actually represents a distribution of windows and is used by our methodology to perform hypothesis testing.
\end{itemize} 

\subsubsection{Constructing a Graph Using CAN Messages}
\label{subsubsec:graph}
In~\cite{Marchetti:2017}, the authors propose an 
IDS based on
recurring sequential message IDs, but this model is vulnerable to intelligent 
attacks. We consider this a starting point for our methodology and 
incorporate graph theory to build a solid IDS platform. 
We divide the CAN bus messages into a number of windows and then 
try to derive the relationships among all the arbitration IDs for each window.  

Empirically, graphs are a common method to indicate the relationships among data. 
Their purpose is to present data that are too complicated to express using simple 
text or other forms of data structure. For that reason, they have now turned into 
one of the most popular fields of research. As graph theory can represent 
complex relationships of data in a very simple manner, we leverage graph 
data structure to represent CAN bus data windows in a meaningful structure.

For any given raw CAN dataset, the proposed Algorithm~\ref{alg:graph_building} 
constructs graphs for every 200 messages (to which we refer as a window) and finally 
returns the overall constructed graph lists. From Line~\ref{alg1:line_ini} 
to Line~\ref{alg1:line_gpin2}, all the necessary variables are initialized. 
Then we compute the total number of messages in the given CAN dataset in Line~\ref{alg1:msgSize}.
In Line~\ref{alg1:forloopst},  a loop is used to iterate over every CAN bus 
message from the CAN bus dataset. As our methodology considers the arbitration 
ID as the node of a graph, so we need to extract the arbitration ID from 
each CAN message in the dataset. From Line~\ref{alg1:loop1} to Line~\ref{alg1:loop4}, 
we extract the adjacent CAN messages and their corresponding IDs. Then the algorithm 
constructs an adjacency list from the arbitration IDs 
extracted from two sequential CAN messages in Line~\ref{alg1:loop5}. 

\begin{algorithm}
	\begin{small}
		\caption{Graph building algorithm}
		\label{alg:graph_building}
		\begin{algorithmic}[1]
			\State {\bf Input:} $CANMessageList[Msg_1, Msg_2,\dots,Msg_n]$ \Comment{Raw CAN bus data array} \\
			{\bf Output:} $GrapList[GP_1, GP_2,\dots, GP_n]$ \Comment{Graph array of CAN bus data}
			\State 
			\State {\bf Initialize:} $GraphList  \leftarrow [$ $]$ \label{alg1:line_ini}
			\State $PreviousID \leftarrow -1$ \label{alg1:line_idin}
			\State $CurrentGraph \leftarrow \{ \}$ \label{alg1:line_gpin} \Comment{Start with an empty graph}
			\State $adjacencyList \leftarrow \{ \}$ \label{alg1:line_adjin} \Comment{A dictionary for adjacency list}
			\State $graphCount \leftarrow 1$ \label{alg1:line_gpin2} 
			\State $N \leftarrow length(CMsgList[Msg_1, Msg_2,\dots,Msg_n])$ \label{alg1:msgSize}
			\For {\texttt{$index$ $in$ $range$ $(0, N-1)$}} \label{alg1:forloopst} \Comment{Loop through all the CAN messages}
			\State      $CANSingleMsg1 \leftarrow CMsgList[index]$ \label{alg1:loop1}
			\State      $CANSingleMsg2 \leftarrow CMsgList[index + 1]$ \label{alg1:loop2}
			\State      $arbitrationID1 \leftarrow ExtractID(CANSingleMsg1)$ \label{alg1:loop3} \Comment{Extraction of Arbitration ID from raw CAN data}
			\State      $arbitrationID2 \leftarrow ExtractID(CANSingleMsg2)$ \label{alg1:loop4} 
			\State $adjacencyList \leftarrow linkGraphNodes(CANSingleMsg1, CANSingleMsg2)$ \label{alg1:loop5} \Comment{Create link between the two graph Nodes}
			\If{($length(adjacencyList) == 200$}  \label{alg1:loop11} \Comment{If it true then
				adjacencyList is a graph built from 200 CAN messages}
			\State $nodeNumber \leftarrow countNodeNumber (adjecencyList)$ \label{alg1:alg1Line16} \Comment{Count number of nodes in the current graph}
			\State $edgeNumber \leftarrow countEdgeNumber (adjecencyList)$ \label{alg1:alg1Line17} \Comment{Count number of edges in the current graph}
			\State $Maxdegree \leftarrow countDegree (adjecencyList)$ \label{alg1:alg1Line18} \Comment{Count maximum degree of each ID from the current graph}
			\State $currentGraph \leftarrow fetchGraphProperties (adjacencyList)$ \label{alg1:Line19} 
			\State $adjecencyList \leftarrow \{ \}$  \label{alg1:alg1Line20}
			\State $currentGraph \leftarrow \{ \}$  \label{alg1:alg1Line21}
			\EndIf \label{alg1:alg1Line22}
			\EndFor \label{alg1:endforloop}
			\State $return$ $GraphList$ \label{alg1:return}
		\end{algorithmic}
	\end{small}
\end{algorithm}

\subsubsection{Extracting Graph-Based Features}
\label{subsubsec:graph_features}    
The graph data structure has several basic properties like the number of 
edges, the number of nodes, the in-degree, the out-degree, etc. In this step, 
our proposed method will characterize each of the message windows and then will build 
the population window by extracting graph properties. In order to extract graph 
properties from a graph constructed using a window of CAN messages, our methodology 
uses the outcome of step 1 of the proposed Algorithm~\ref{alg:graph_building}. 
In Algorithm~\ref{alg:graph_building}, Line~\ref{alg1:alg1Line16} to Line~\ref{alg1:alg1Line18} 
extracts the node number, edge number, and maximum degree from the single constructed graph and 
stores them in the list of graph properties in Line~\ref{alg1:Line19} using the method fetchGraphProperties(). 
After iterating through 200 messages, we initialize the adjacency list and the current 
graph in Line~\ref{alg1:alg1Line20} and Line~\ref{alg1:alg1Line21}, respectively.
In Line~\ref{alg1:return}, 
we return the whole graph list, which is the population window in our methodology.  

\subsubsection{Proposed Hypothesis Based on Safe Population Window}
\label{subsubsec:hypothesis}

In this step of our proposed intrusion detection methodology, we try to build a 
hypothesis based on the information of the population window. We have used the popular 
chi-squared statistical test to build the hypothesis~\cite{Cochran:1952}. 
Using this statistical analysis, we compute a threshold value. The threshold 
value helps us in the next step to detect the anomalous population. 
Besides, we introduce a conventional median test to detect a strong replay attack.   

Algorithm~\ref{alg:attck_detect} represents the chi-squared test on our 
population window. It takes two lists of graphs as inputs and then outputs a 
boolean value. Out of the two inputs, one is the attack-free graph population 
and the other is the graph population under test. The boolean value in the 
output represents whether there is any attack happening or not. In between 
Line~\ref{alg2:line_line7} to Line~\ref{alg2:line_line6} we initialize 
the variables that are needed for our proposed methodology. After that, 
the algorithm extracts the edges of each graph from the graph list from 
Line~\ref{alg2:line11} to Line~\ref{alg2:line13} for safe attack-free 
distribution. Finally, a hypothesis is constructed for the safe edge distribution in Line~\ref{alg2:line17}.

\begin{algorithm}
	\begin{small}
		\caption{Proposed attack detection algorithm}
		\label{alg:attck_detect}
		\begin{algorithmic}[1]
			\State {\bf Input:} $GrapList_{Base}[GP_1, GP_2,\dots, GP_n]$ \Comment{Graph array of attack free CAN bus data},  $GrapList_{Test}[GP_1, GP_2,\dots, GP_n]$ \Comment{Graph array of test data}\\
			{\bf Output:} $isAttacked$ \Comment{True if the input graph is attacked or false otherwise}
			\State
			\State {\bf Initialize:} $ edgeList_{Base}\leftarrow [$ $]$ \label{alg2:line_line7}
			\State $ edgeList_{Test}\leftarrow [$ $]$ \label{alg2:line_line8}
			\State $Chi_{Null} \leftarrow 0$ \label{alg2:line_line9}
			\State $Chi_{Test} \leftarrow 0$ \label{alg2:line_line10}
			\State $Median_{Null} \leftarrow 0$ \label{alg2:line_line10_1_00}
			\State $Median_{Test} \leftarrow 0$ \label{alg2:line_line10_1_01}
			\State $\sigma_{Null} \leftarrow 0$ \label{alg2:line_line10_1_02}
			\State $threshold \leftarrow 0$ \label{alg2:line_line10_1}
			\State $N_{Base} \leftarrow length(GrapList_{Base}[GP_1, GP_2,\dots, GP_n])$ \label{alg2:line_line5} 
			\State $N_{Test} \leftarrow length(GrapList_{Test}[GP_1, GP_2,\dots, GP_n])$ \label{alg2:line_line6} 
			\For {\texttt{$index$ $in$ $range$ $(0, N_{Base})$}} \label{alg2:line11}
			\State      $edgeList_{Base} \leftarrow fetchEdgeNumbers(GraphList_{Base}[index])$ \label{alg2:line_line12}
			\EndFor \label{alg2:line13}
			\For {\texttt{$index$ $in$ $range$ $(0, N_{Test})$}} \label{alg2:line14}
			\State      $edgeList_{Test} \leftarrow fetchEdgeNumbers(GraphList_{Test}[index])$ \label{alg2:line15}
			\EndFor \label{alg2:line16}
			\State      $Chi_{Null} \leftarrow ExtractDistribution(edgeList_{Base})$ \label{alg2:line17} \Comment{Construct hypothesis using attack free graph}
			\State      $Chi_{Test} \leftarrow ExtractDistribution(edgeList_{Test})$ \label{alg2:line18} \Comment{Construct hypothesis using test data}
			\State      $threshold \leftarrow FindSignificanceLevel(Chi_{Null})$ \label{alg2:line19} \Comment{Find threshold using the base distribution}
			\State      $\{Median_{Null},\sigma_{Null}\} \leftarrow ExtractDistribution(edgeList_{Base})$ \label{alg2:line19_01} \Comment{Compute median and standard deviations using attack free graphs}
			\State      $Median_{Test} \leftarrow ExtractDistribution(edgeList_{Test})$ \label{alg2:line19_02} \Comment{Compute median using test data}
			\If{($Chi_{Test} \leq  threshold$)}  \label{alg2:line20}
			\State $isAttacked \leftarrow True$ \Comment{Attack detected} \label{alg2:line21}
			\ElsIf{($Median_{Test} >  (Median_{Null} + 3 \sigma_{Null}$))}  \label{alg2:line21_01}
			\State $isAttacked \leftarrow True$ \Comment{Attack detected} \label{alg2:line21_02}
			\Else \label{alg2:line22}
			\State $isAttacked \leftarrow False$ \label{alg2:line23} \Comment{CAN data is safe} 
			\EndIf \label{alg2:line24}
			\State $return$ $isAttacked$ \label{alg2:line25}
		\end{algorithmic}
	\end{small}
\end{algorithm}

\subsubsection{Comparing Test Population Window to the Base Population Window}
\label{subsubsec:comparing}
This step of our proposed intrusion detection methodology consists of two functionalities. 
The first one is to calculate the chi-square value for the test population 
window and then compare the value with the threshold. 
The latter one is to calculate the median value for the 
test population window and then compare the value with the outlier.
By using Equation~\ref{eq:chi_test1} 
and Equation~\ref{eq:chi_test2} we can detect the anomalous population.    
\begin{equation}
\label{eq:chi_test1}
Chi_{Test} <= threshold, [Chi_{Test} = X_c^2]; No \quad attack \\
\vspace{-0.8cm}
\end{equation}
\begin{equation}
\label{eq:chi_test2}
Chi_{Test} >  threshold, [Chi_{Test} = X_c^2]; Attack
\end{equation}
In Line~\ref{alg2:line14} to Line~\ref{alg2:line16} of Algorithm~\ref{alg:attck_detect}, 
we extract the edges from the list of graphs that will be tested. In  Line~\ref{alg2:line18}, 
a test hypothesis is made based on the same rules as the attack-free distribution 
hypothesis or the base hypothesis and the threshold is defined in Line~\ref{alg2:line19} 
based on the details of the base hypothesis. 
To detect the replay attack, we incorporate the median test~\cite{Siegel:1988}. For this, the 
Algorithm~\ref{alg:attck_detect} compute median ($Median_{Null}$) and standard 
deviation $\sigma_{Null}$ using attack free data and $Median_{Test}$ of attacked 
data in Line~\ref{alg2:line19_01} to Line~\ref{alg2:line19_02}. 
Finally, the algorithm takes the decision about 
the test distribution by comparing the test hypothesis with the threshold or 
our defined outliers ($Median_{Null} + 3 \sigma_{Null})$ from 
Line~\ref{alg2:line20} to Line~\ref{alg2:line24} and returns 
whether the CAN data is safe or not in Line~\ref{alg2:line25}. 
Our assumed outliers exhibits excellent results and we will discuss 
in detail in Section~\ref{subsec:exp_data_ana} and Section~\ref{subsec:exp_DoS}.  

\subsection{Exploratory Data Analysis}
\label{subsec:exp_data_ana}
We have chosen a real vehicle dataset provided by the Hacking and 
Countermeasure Research Lab~\cite{Lee_data:2018}. The dataset includes both 
attack-free and corrupted data with various kinds of attacks. 
Those attacks include DoS, fuzzy, spoofing, and replay attacks on CAN data.

\begin{figure}[t!]
	\centering
	\vspace{-0.0cm}
	\includegraphics[width = 0.35\textwidth]{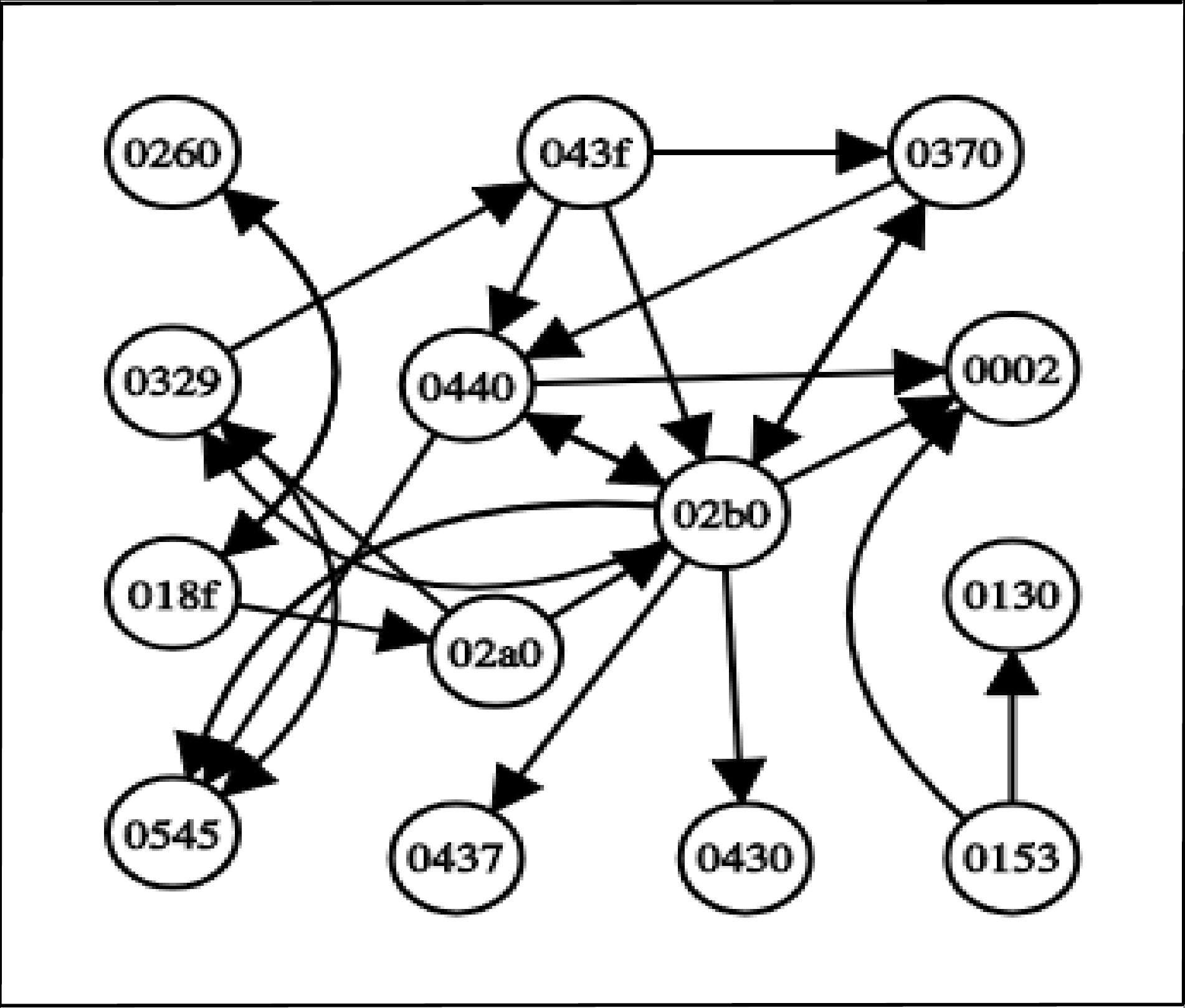}
	\caption {We use real CAN message data to build a directed graph that represents the sequence of messages.}
	\label{fig:graph}
	\vspace{-0.0cm}
\end{figure}

First, we build graphs using Algorithm~\ref{alg:graph_building} with raw CAN bus data. 
Figure~\ref{fig:graph} shows an example of a graph generated using the proposed methodology. 
The nodes of the graph represent arbitration IDs of the CAN bus, and the edge between two 
nodes indicates CAN bus sequential messages. The direction of the edge indicates the 
order of the sequence of the messages. For example, if node $043f$ has an edge 
with node $0440$ and the direction of the edge goes from $043f$ to $0440$, it 
means arbitration ID $0440$ was followed by the message with arbitration ID $043f$. 

\begin{figure}
    \centering
    \vspace{-0.5cm}
    \subfigure[]{\includegraphics[width=0.245\textwidth]{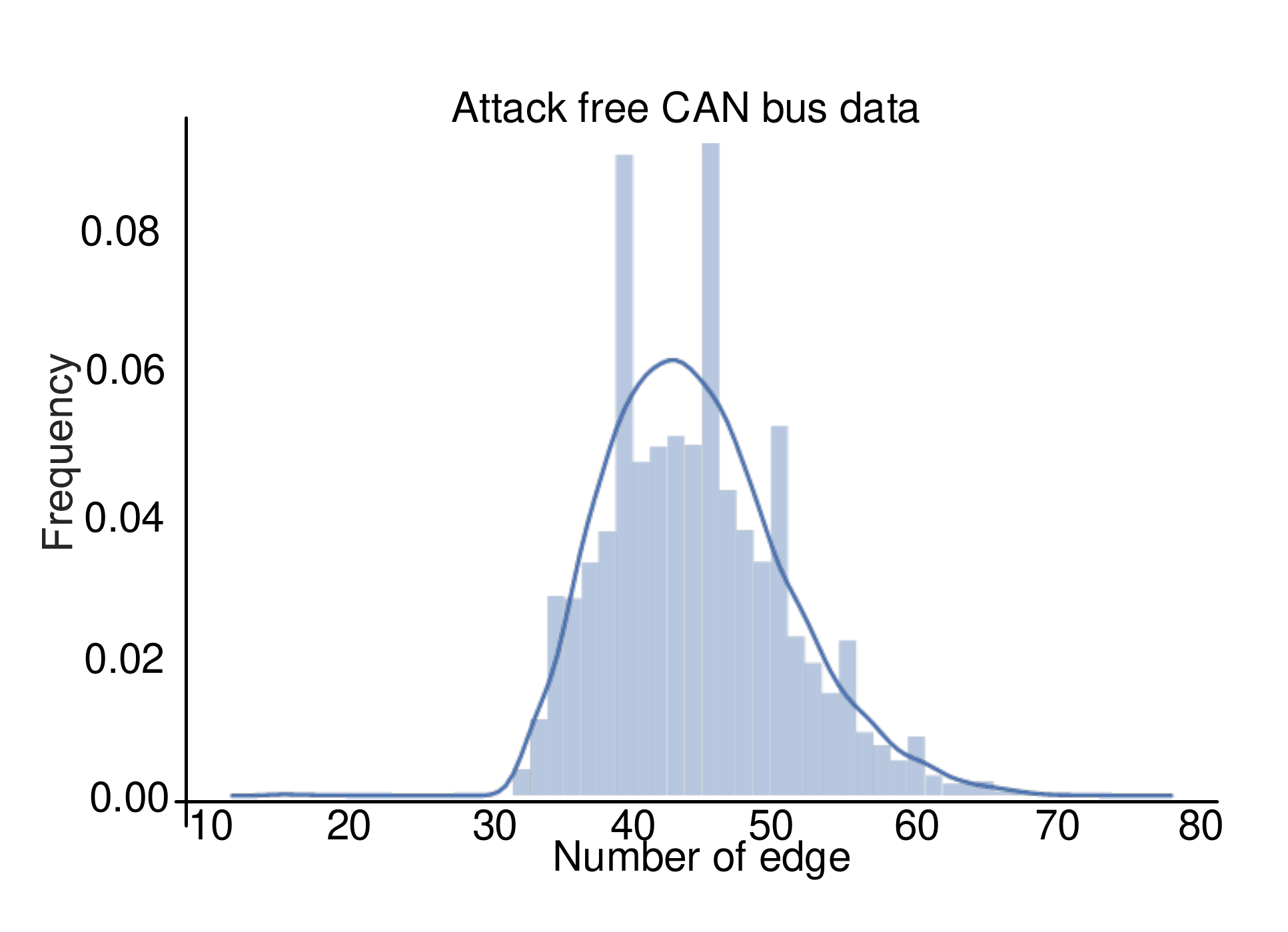}} 
    \subfigure[]{\includegraphics[width=0.225\textwidth]{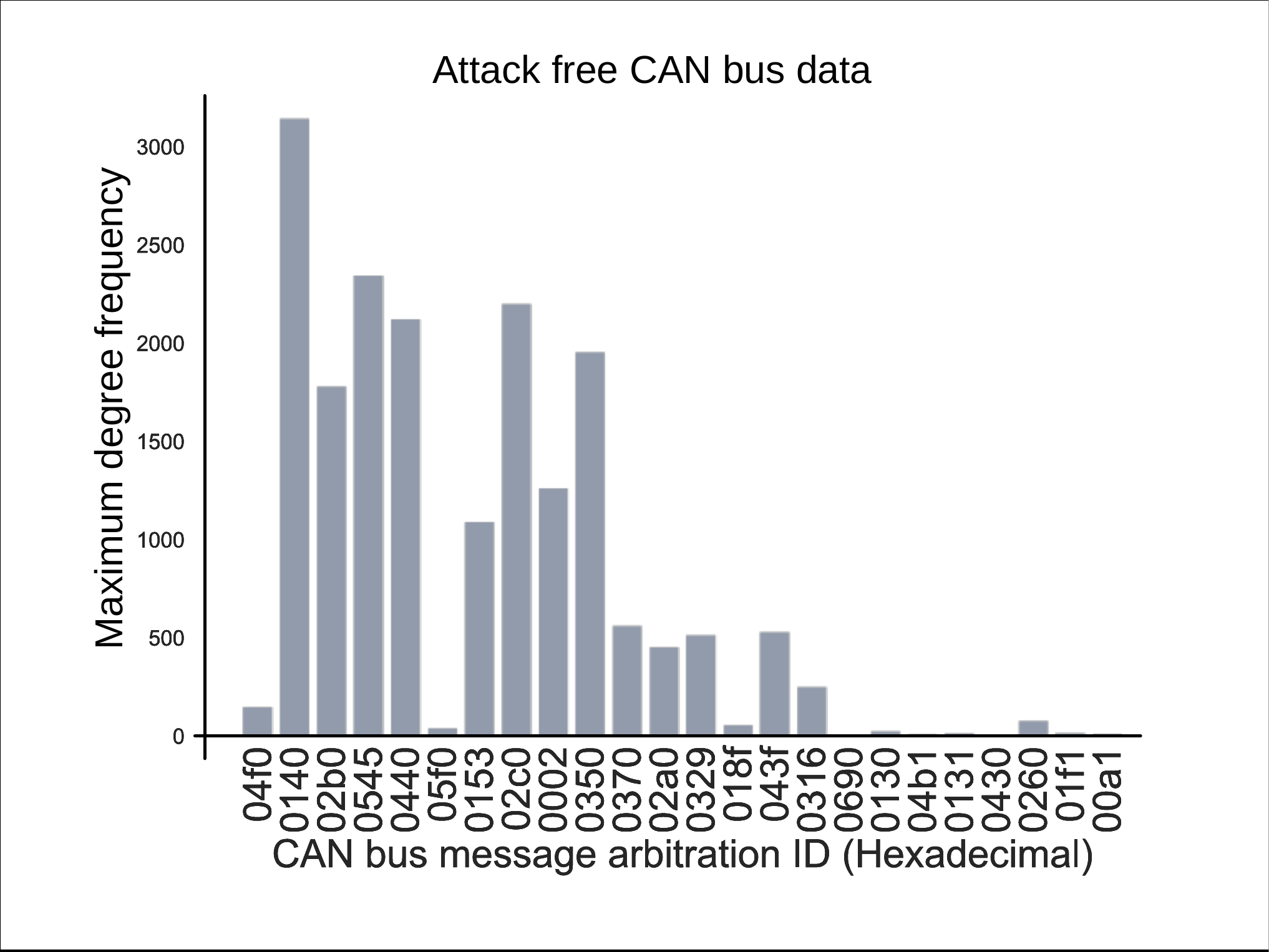}} 
    \caption{(a) Attack-free CAN data edge distribution shows a normal distribution, and (b) the attack-free CAN data maximum degree distribution exhibits a regular pattern.}
    \label{fig:attackfree_2fig}
    \vspace{-0.5cm}
\end{figure}

First, we divided the real vehicular CAN bus data into a few windows. 
The size of each window is 200 messages. However, this number is user-defined, 
and it is possible to change depending on the design robustness. 
At a 1Mbit/sec speed, a CAN transmits about 8.7K messages/sec. Hence, designers 
can select how frequently they want to authenticate the CAN. 
After selecting the window size, we build the graph for each of the windows 
and derive the common graph properties like edge number and maximum 
degree distribution for each window. The 
attack-free CAN data edge distribution shows a normal distribution, as 
shown in Figure~\ref{fig:attackfree_2fig}(a).
Similarly, the attack-free 
CAN data maximum degree distribution exhibits a regular pattern, as 
shown in Figure~\ref{fig:attackfree_2fig}(b).

\begin{figure}
    \centering
    \vspace{-0.5cm}
    \subfigure[]{\includegraphics[width=0.245\textwidth]{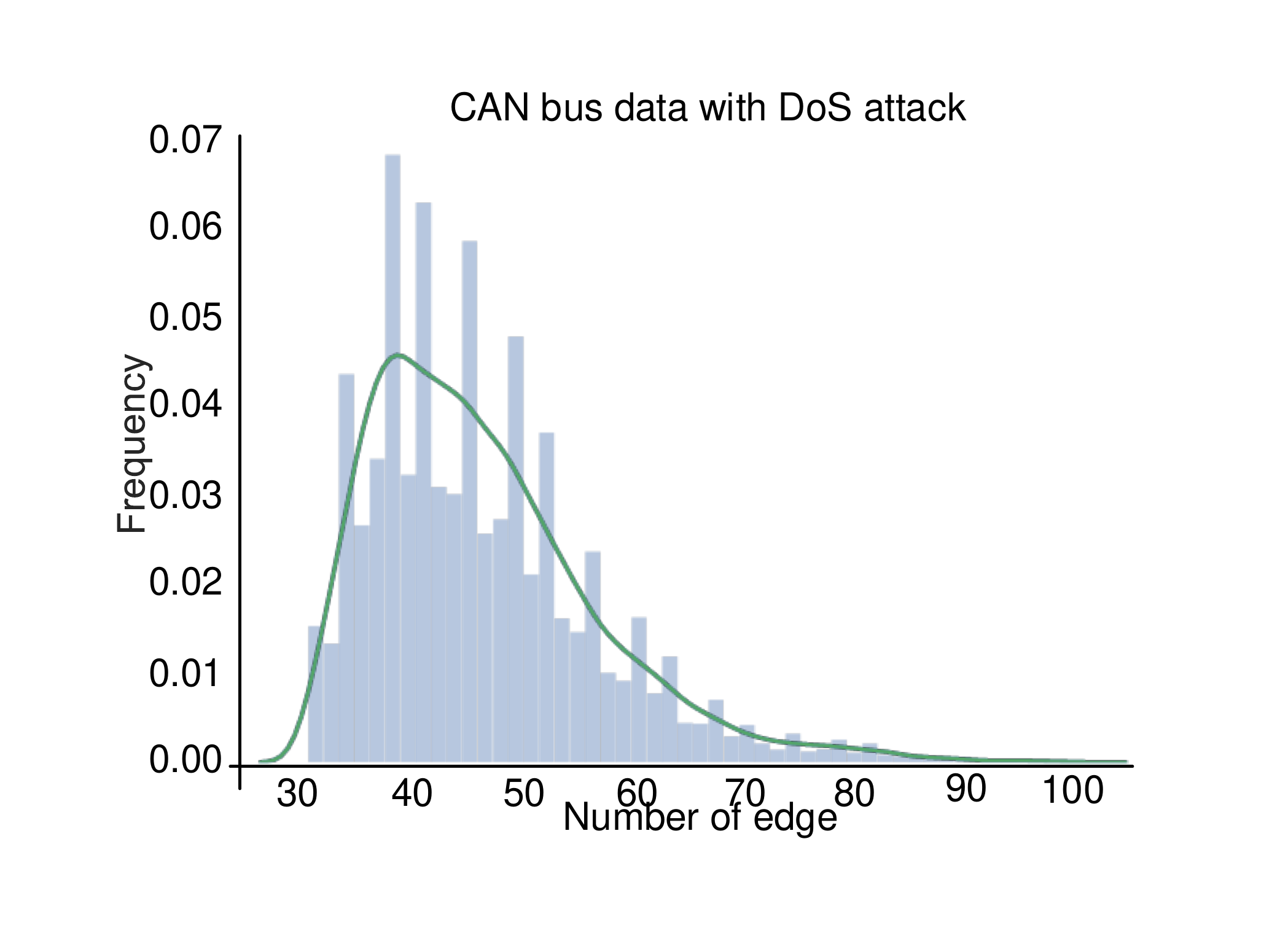}} 
    \subfigure[]{\includegraphics[width=0.225\textwidth]{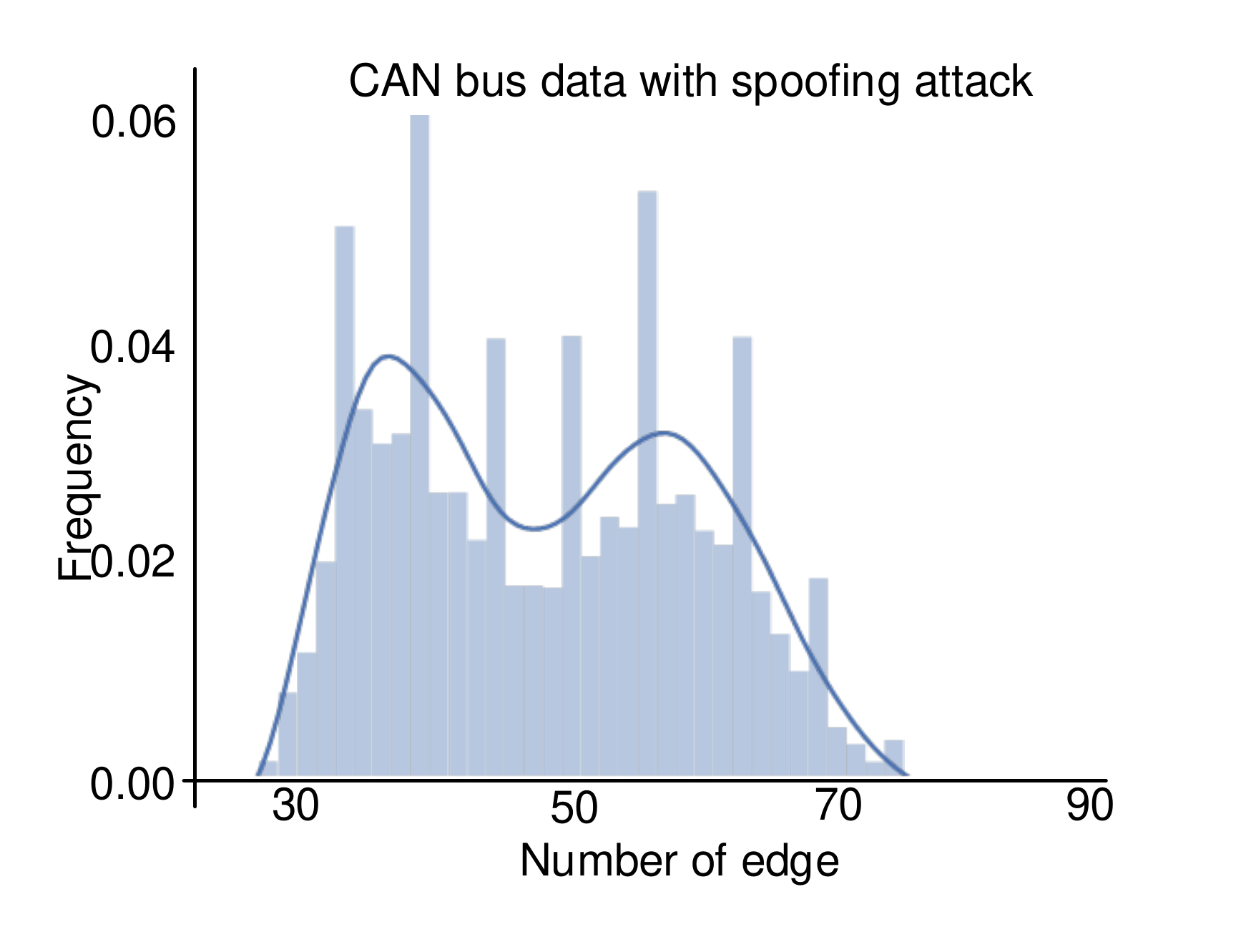}} 
    \subfigure[]{\includegraphics[width=0.244\textwidth]{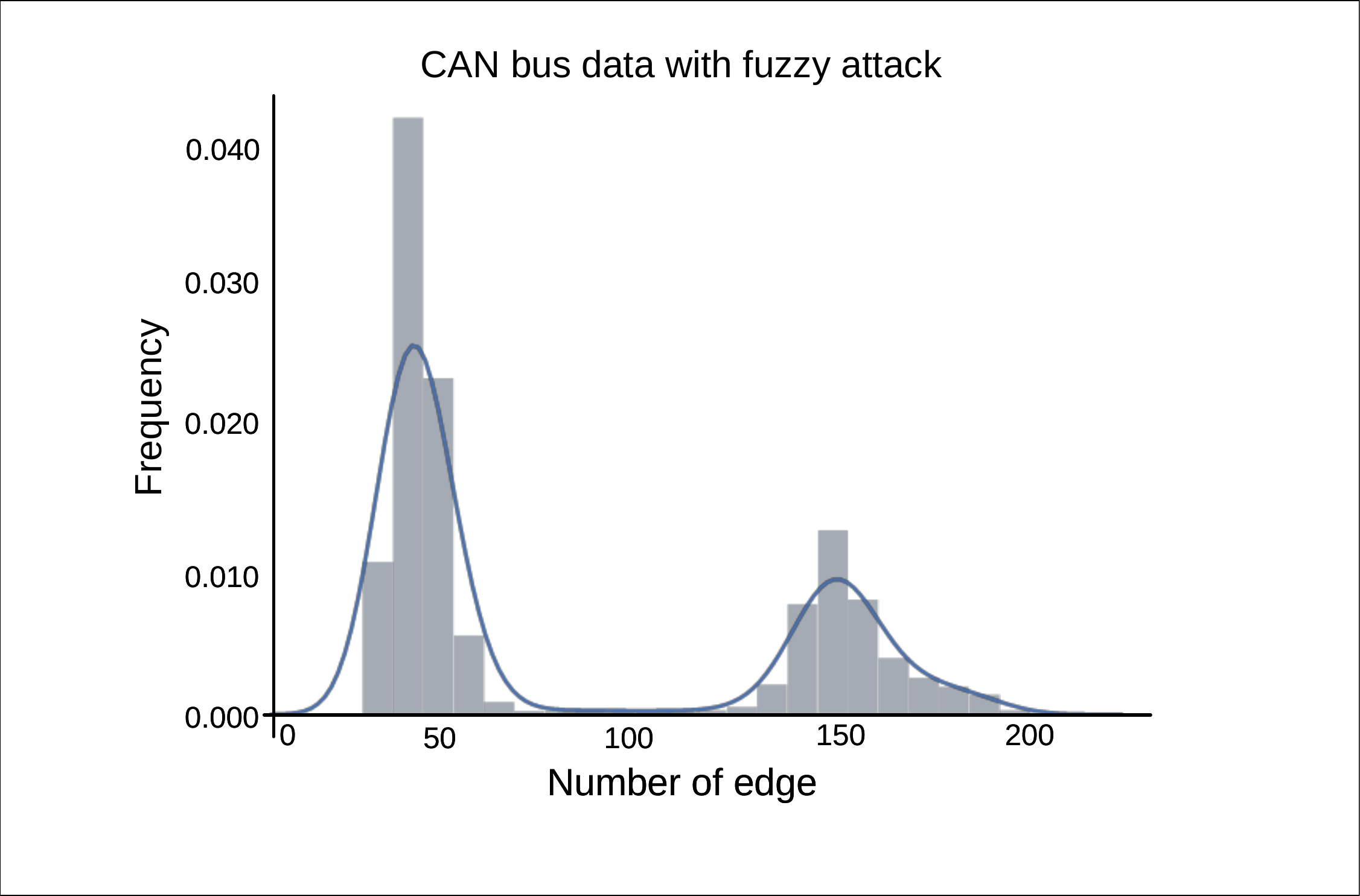}}
    \subfigure[]{\includegraphics[width=0.233\textwidth]{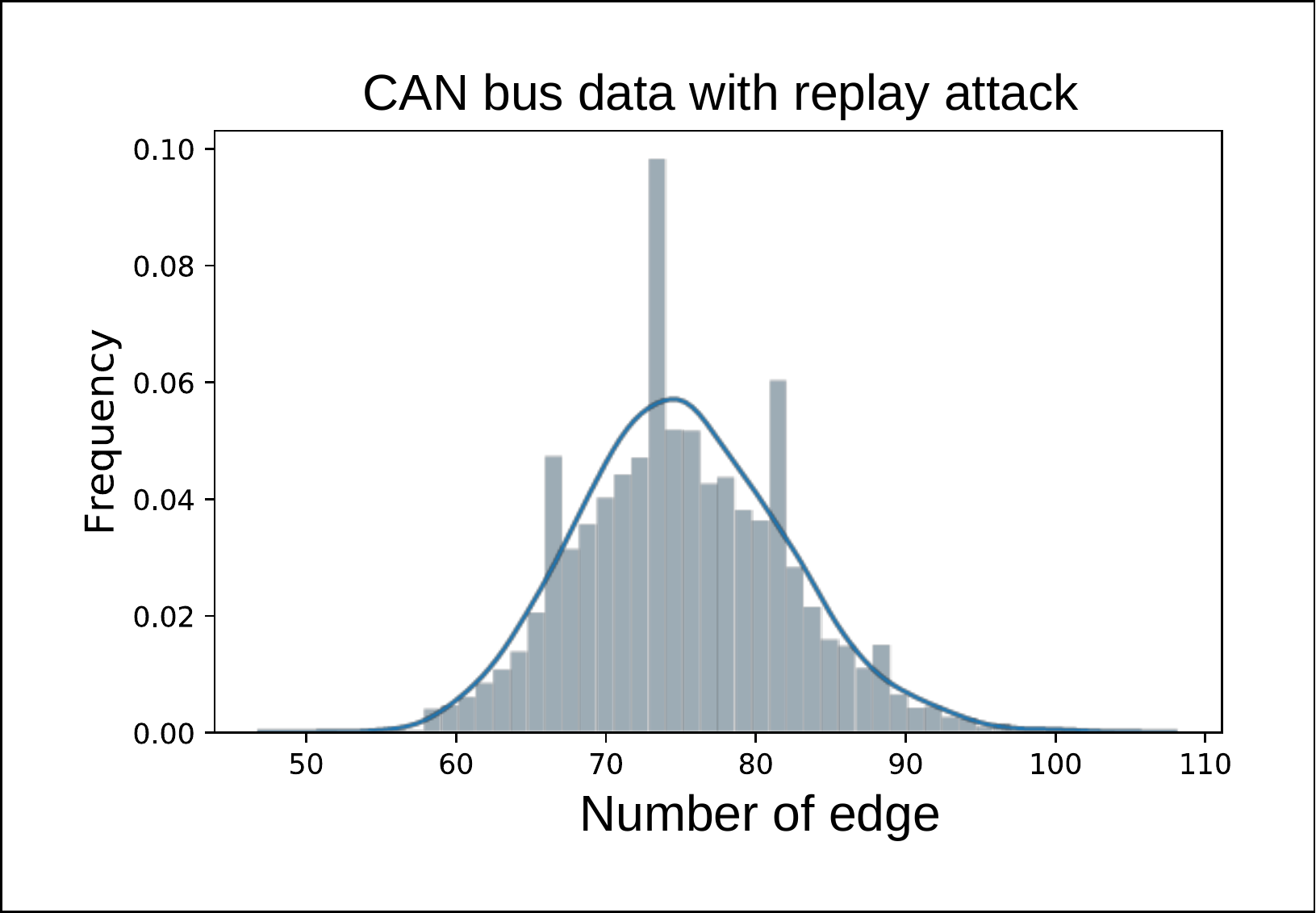}}
    \caption{(a) The DoS-attacked CAN data edge distribution shows a positively skewed distribution, (b) the spoofing-attacked CAN data edge distribution shows a bimodal distribution, (c) similar to spoofing attack, the fuzzy-attacked CAN data edge distribution shows a bimodal distribution, and (d) similar to attack-free CAN data, the edge distribution of replay-attacked data shows a normal distribution.}
    \label{fig:edge_4fig}
    \vspace{-0.5cm}
\end{figure}

\begin{figure}
    \centering
    \vspace{-0.5cm}
    \subfigure[]{\includegraphics[width=0.230\textwidth]{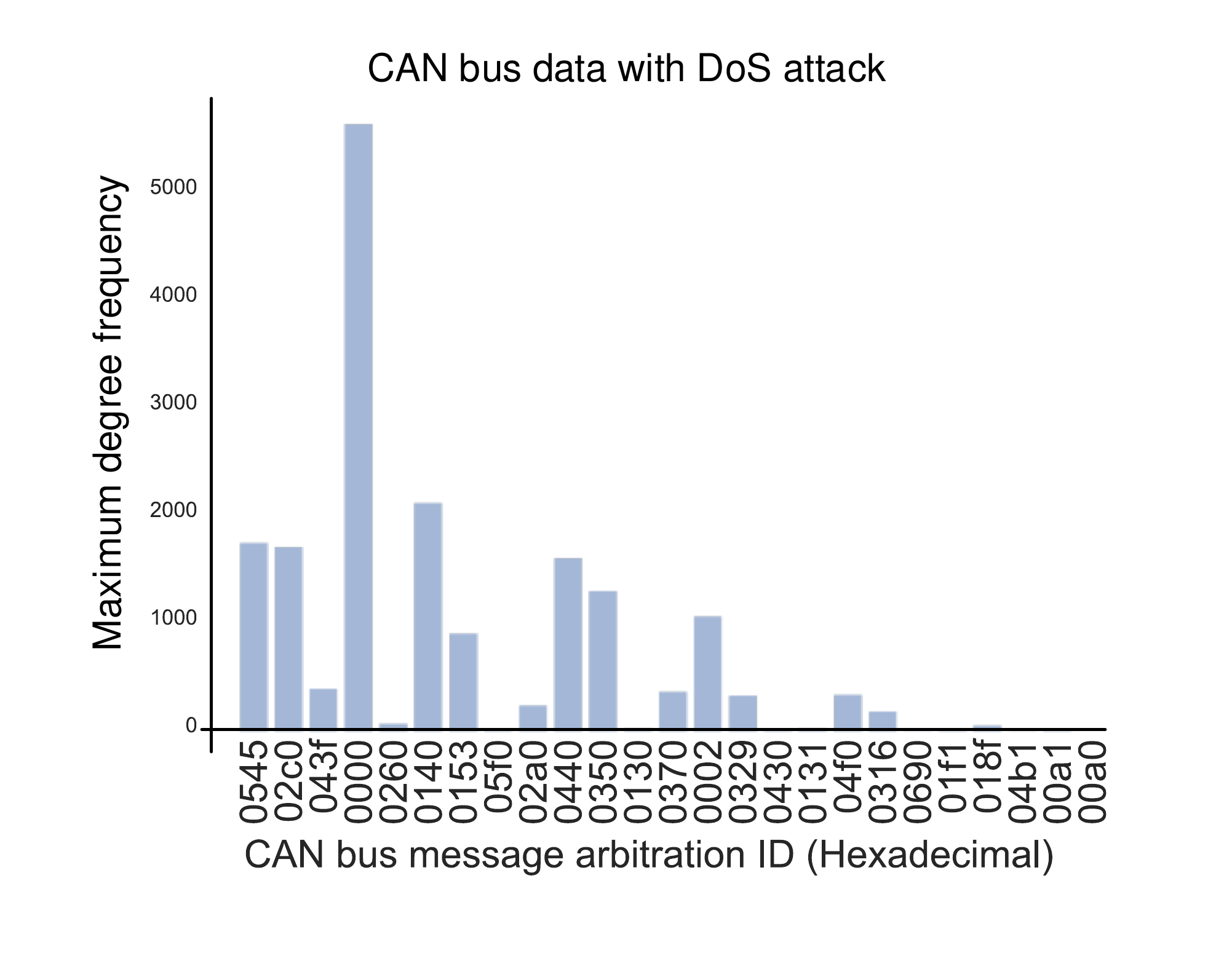}} 
    \subfigure[]{\includegraphics[width=0.225\textwidth]{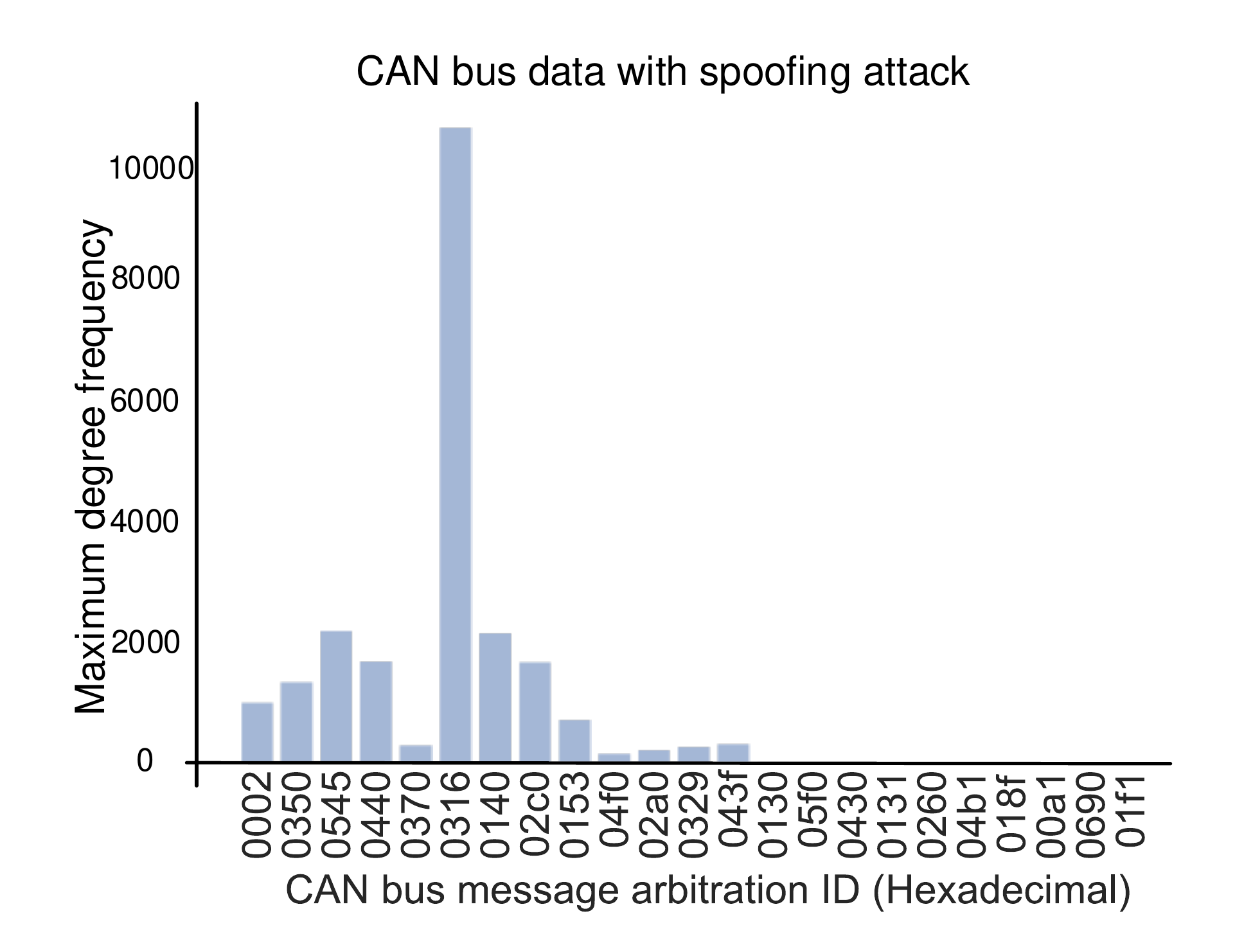}} 
    \subfigure[]{\includegraphics[width=0.244\textwidth]{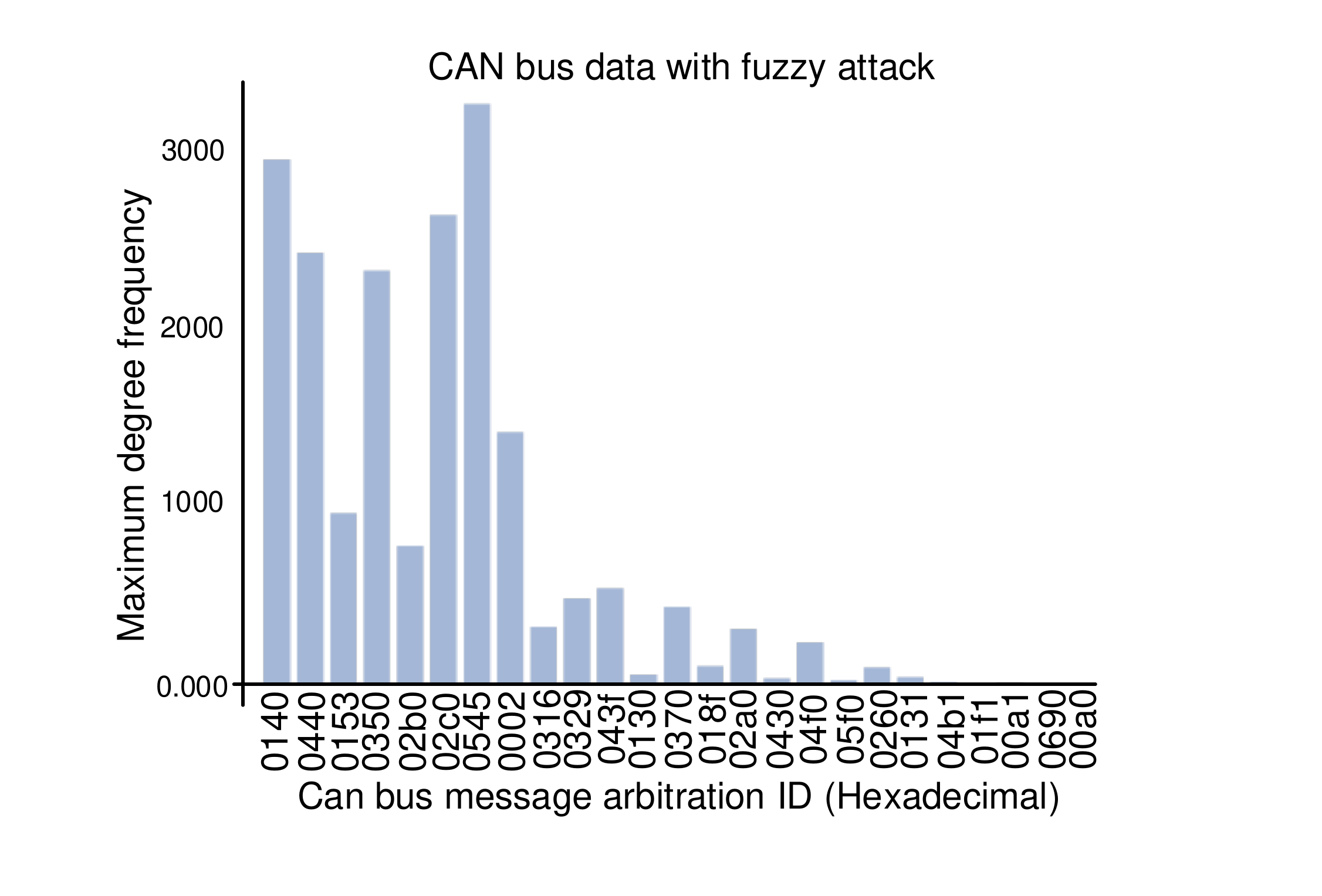}}
    \subfigure[]{\includegraphics[width=0.233\textwidth]{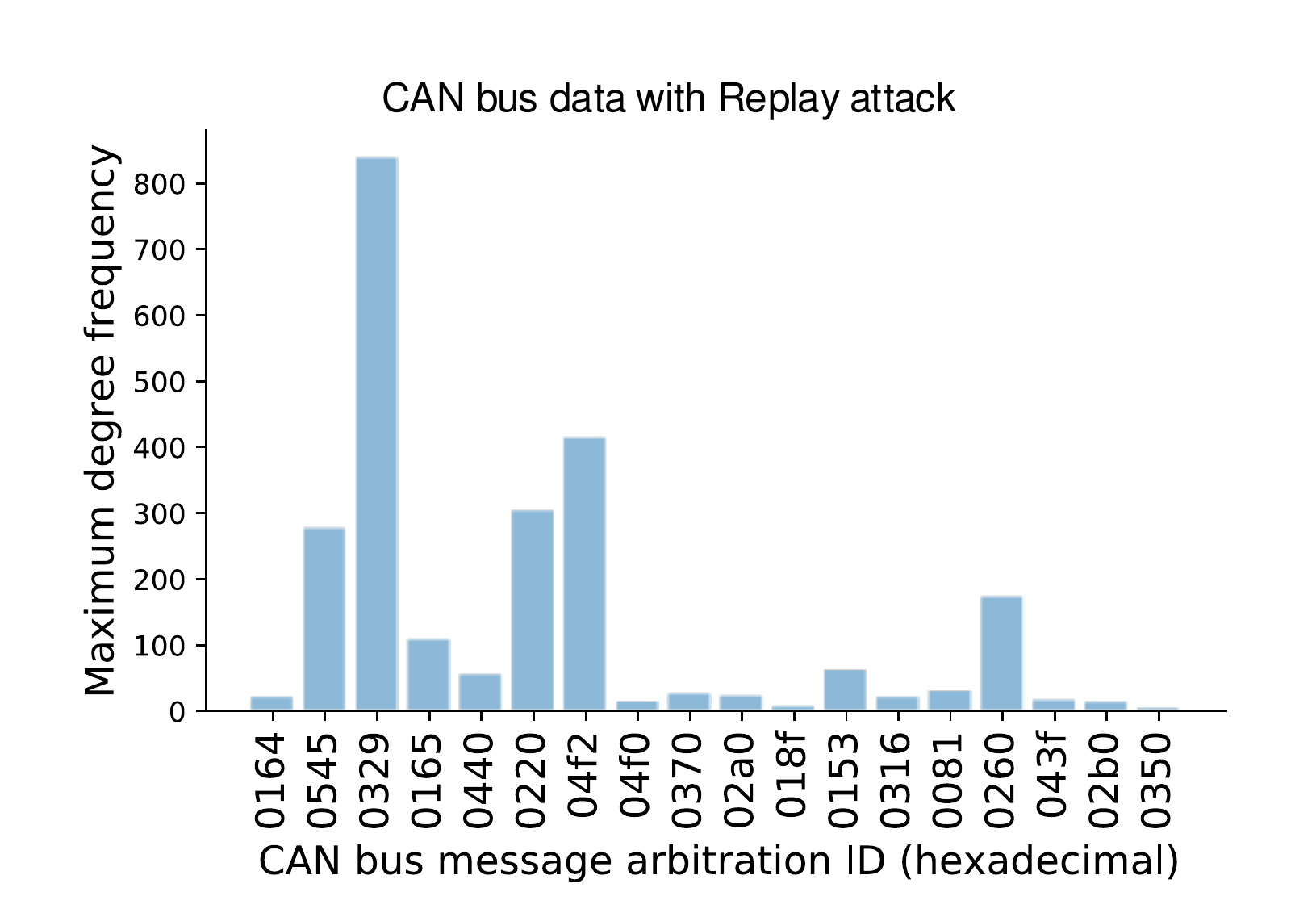}}
    \caption{(a) The DoS-attacked CAN data maximum degree distribution exhibits an irregular pattern with arbitration ID $0000$, (b) the spoofing-attacked CAN data maximum degree distribution exhibits an irregular pattern with arbitration ID $0316$, (c) The fuzzy-attacked CAN data maximum degree distribution exhibits a regular pattern, and (d) similar to the attack-free CAN data, the replay attacked maximum degree distribution exhibits a regular pattern.}
    \label{fig:Maxdegree_4fig}
    \vspace{-0.5cm}
\end{figure}

Now we will discuss the graph properties of the dataset with different kinds of attacks.
Figure~\ref{fig:edge_4fig}(a) represents the distribution of edges for a
DoS attack. Unlike attack-free CAN data, the DoS-attacked data do not 
exhibit a normal distribution. 
Figure~\ref{fig:Maxdegree_4fig}(a) shows the situation of the maximum degree 
for a CAN dataset with a DoS attack. Clearly, a single arbitration ID ($0000$) 
dominates the distribution and occupies the CAN network with the highest-priority messages.

Unlike the DoS attack, the graphs with a spoofing-attacked dataset show 
a bimodal distribution with two distinct peaks, as shown in Figure~\ref{fig:edge_4fig}(b).  
However, similar to a DoS attack, the spoofing-attacked maximum edge distribution 
has a distinguishable high occurrence of ID $0316$ compared to the other IDs, 
as shown in Figure~\ref{fig:Maxdegree_4fig}(b). 

Similar to the spoofing attack, the graphs with a fuzzy-attacked 
dataset show a bimodal distribution with two distinct peaks, as 
shown in Figure~\ref{fig:edge_4fig}(c). However, unlike the spoofing attack, 
the fuzzy-attacked maximum edge distribution has no distinct 
characteristics compared to the attack-free data as shown in Figure~\ref{fig:Maxdegree_4fig}(c).

Unlike other attacks, the graphs with a replay-attacked dataset show a 
normal distribution, as shown in Figure~\ref{fig:edge_4fig}(d). In addition, 
replay-attacked maximum edge distribution has no distinct characteristics 
compared to the attack-free data, as shown in Figure~\ref{fig:Maxdegree_4fig}(d). 

Apart from impersonation or replay attacks, the data distributions of different attacks are not only different from the attack-free CAN
data distribution but also different from each other. 
We summarize in statistical terms the overall situation of the data distributions in Table~\ref{tab:explo_data}.
In the exploratory data analysis, we consider the central tendency or mean of the distribution and the
asymmetry of a probabilistic distribution. In our attack-free dataset graph collection, 
the edge distribution has a mean of 44.6. On the other hand, the DoS,
fuzzy, spoofing, and replay attacks have a mean of 46.6, 49.1, 79.8, and 75.17, respectively.
In addition, we compute the median of edge distribution which is important for 
outliers detection, as shown in Table~\ref{tab:explo_data}. The attack free, DoS,
fuzzy, spoofing, and replay attacks edge distributions have a median of 
44.0, 45.0, 49.0, 46.0, and 75.17, respectively.

Table~\ref{tab:explo_data} also shows the skewness of all the edge distributions for
attack-free and attacked datasets.
Spoofing and replay attack edge distributions
seem symmetric to the attack-free dataset edge distribution.
On the other hand, the attack-free and fuzzy attack edge distributions are
moderately positive-skewed. Finally, the DoS attack edge distribution is highly skewed. If
we look at the maximum degree of CAN arbitration IDs in the graph distribution, 
it shows that different kinds of attack use different IDs, resulting in dissimilar maximum degrees.

Finally, exploratory analysis proves that the conversion between raw CAN data
to the graph gives us a clear indication of an attacked or attack-free CAN bus. 
Using this technique, we can fetch some extraordinary
information from the converted graph. Finally, the graph properties can be used to
distinguish different attacked and attack-free situations of the CAN
bus system.

\begin{table} [t!] \large 
	\renewcommand{\arraystretch}{1.5}
	\caption{In summary, we can say that among all the attacks, impersonation or replay attack is difficult to detect due to the symmetric edge and maximum degree distributions compared to the attack-free data.}
	\label{tab:explo_data}
	\centering
	\scalebox{0.68}{
		{\begin{tabular}{|c|c|c|c|c|c|}
				\hline
				Analysis                & Attack free       & DoS       & Spoofing   & Fuzzy & Replay\\
				\hline
				Mean (edge)             & 44.6              &  46.6     & 49.1    & 79.8 & 75.2 \\
				\hline
				Median (edge)           & 44.0              &  45.0     & 49.0    & 46.0 & 75.0 \\
				\hline
				Skewness (edge)         & Moderate          & High    & Similar   & Moderate &  Similar \\
				\hline
				\multirow{2}{*}{Max degree ID}           & \multirow{2}{*}{$0140$}   &  \multirow{2}{*}{$0000$}    & \multirow{2}{*}{$0316$ }     & \multirow{2}{*}{$0545$ } & \multirow{2}{*}{ $0164$ } \\
				(\%) & (16.9) &  (30.8) &  (46.4) & (16.8) & (31) \\
				\hline
	\end{tabular}}}
	\vspace{-0.20cm}
\end{table}

%% file: analysis.tex
\section{Experimental Results and Evaluation}
\label{sec:analysis}

In order to verify the proposed methodology, we use a real CAN dataset and 
performed analysis on an Intel Xeon(R) 3.8 GHz 8-core processor with 32 GB 
RAM using our proposed algorithm in Python language. 
For our analysis, we consider about 23K graphs. We divide the result 
section into three parts: In Section~\ref{subsec:exp_DoS}, we first will 
discuss the detection methodology for different attacks using the 
proposed graph-based chi-square test and median test.
Then, in Section~\ref{subsec:exp_signifi}, 
we will identify the level of significance (LoS).

\subsection{Attack Detection}
\label{subsec:exp_DoS}
\begin{figure*}[t!]
	\vspace{-0.0cm}
	\includegraphics[width = 1.0\textwidth]{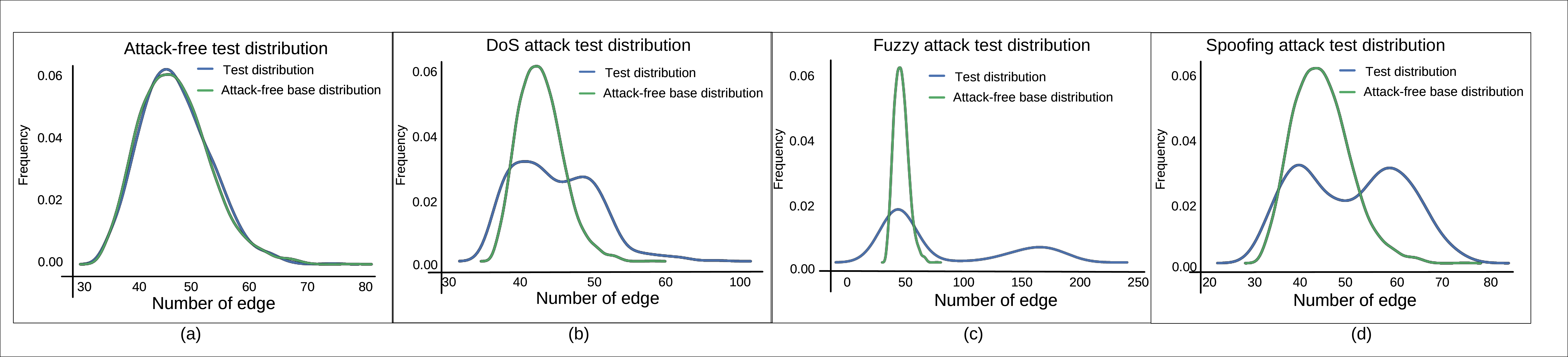}
	\caption {We used a chi-squared test and built a base-hypothesis using attack-free data: (a) expectedly, the attack-free test distribution and base distributions are identical; (b) chi-squared test easily distinguishes the base normal distribution and bimodal distribution of DoS-attacked data; (c) chi-squared test detects the fuzzy attack compared with the base hypothesis; and (d) chi-squared test easily distinguishes the base normal distribution and bimodal distribution of spoofing-attacked data.}
	\label{fig:chi_all}
	\vspace{-0.2cm}
\end{figure*}

For detecting an attack using
the chi-squared test, we build a base hypothesis using the exploratory attack-free CAN
data. We define it as a base distribution. Then, any distribution can be compared
with the base hypothesis and differences can be found easily. Figure~\ref{fig:chi_all}(a) is a 
visual representation of our chi-squared test on an attack-free distribution. The
distribution colored as green represents the safe distribution, and on the other
hand, the blue distribution represents the distribution under the test. 
According to our analysis, any attack-free test data exhibit a similar 
pattern to our base hypothesis, as shown in Figure~\ref{fig:chi_all}(a). 
After that, we built test distributions using DoS-, fuzzy-, spoofing, and
replay-attacked datasets. Figure~\ref{fig:chi_all}(b), Figure~\ref{fig:chi_all}(c), 
and Figure~\ref{fig:chi_all}(d) show the chi-squared test on DoS-, fuzzy-, and 
spoofing-attacked distributions compared with our base hypothesis, respectively.

Among all the CAN monitoring-based attacks, the replay-attacked edge 
distribution (i.e., Figure~\ref{fig:edge_4fig}(d)) shows no difference 
from the attack-free distribution (i.e., Figure~\ref{fig:attackfree_2fig}(a)).
As a result, the chi-squared test can only achieve up to 66\% accuracy. 
Because of this issue, we incorporate the median test by defining outliers, 
considering median and $3\times$ (standard deviation) values. Our prediction 
accuracy increases significantly using this technique.
Clearly, we can easily detect any of those CAN-monitoring-based 
attacks using the proposed methodology. 
For this analysis, we consider only edge distributions.

\begin{figure}
    \centering
    \vspace{-0.5cm}
    \subfigure[]{\includegraphics[width=0.24\textwidth]{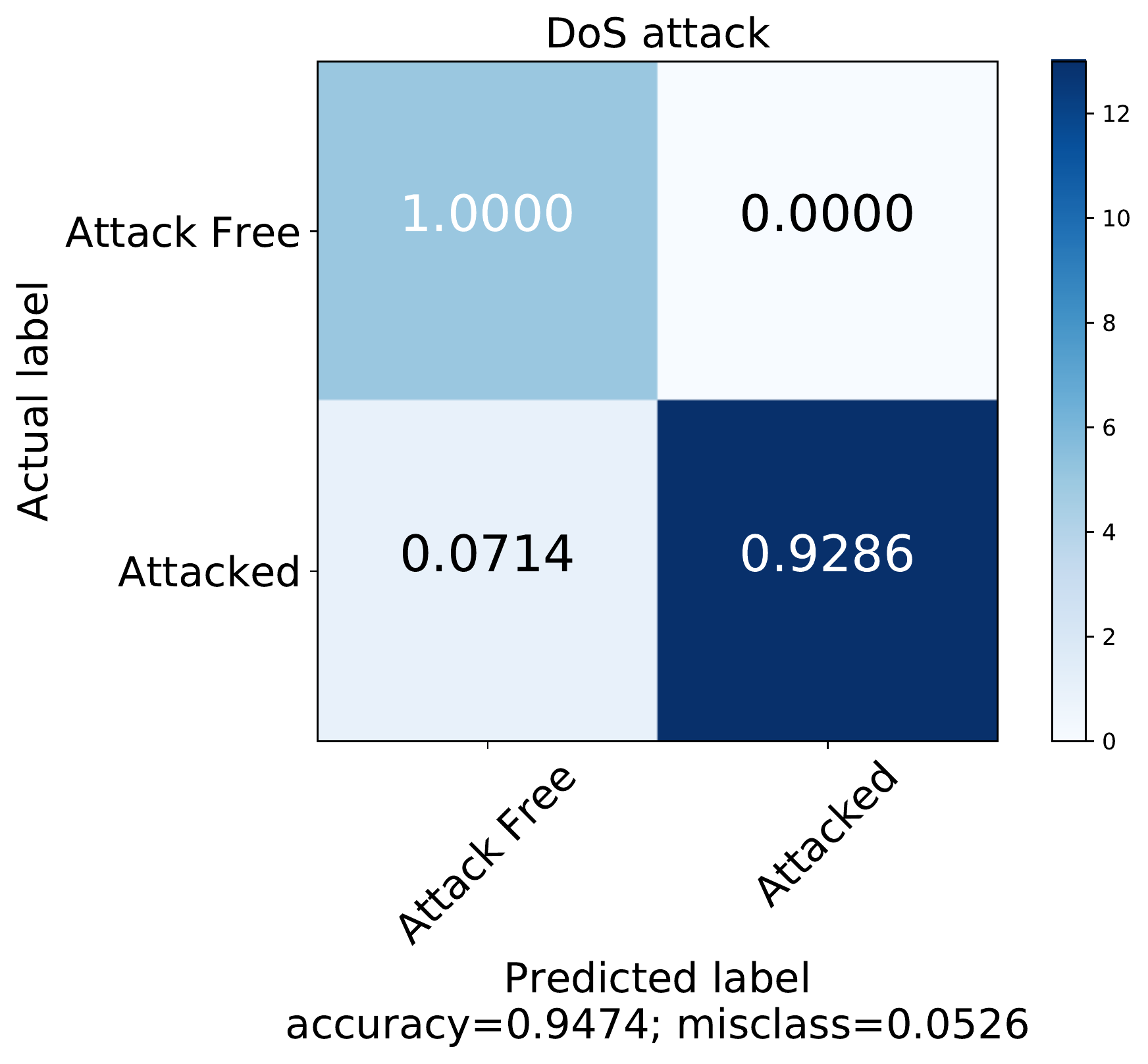}} 
    \subfigure[]{\includegraphics[width=0.24\textwidth]{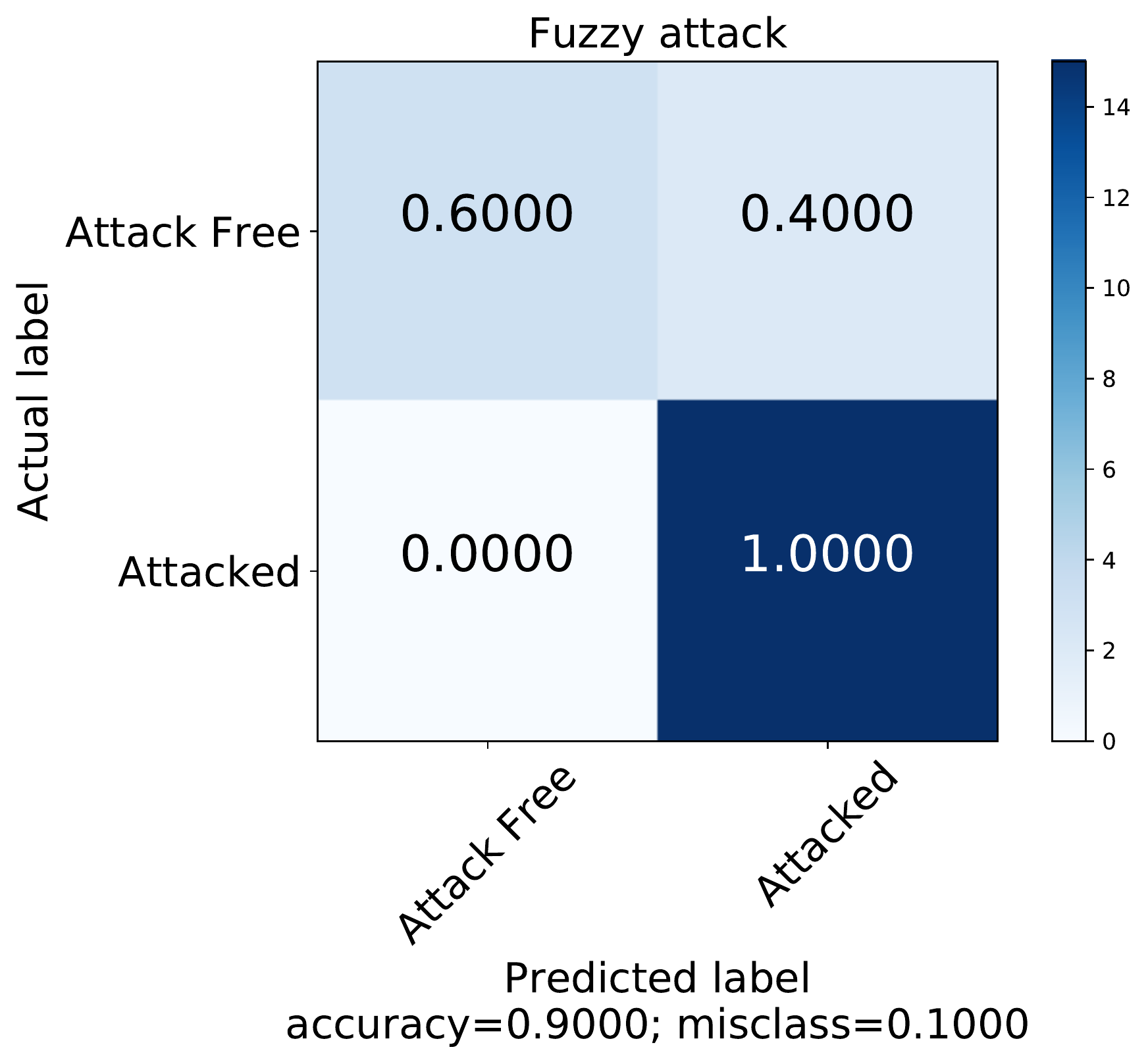}} 
    \subfigure[]{\includegraphics[width=0.24\textwidth]{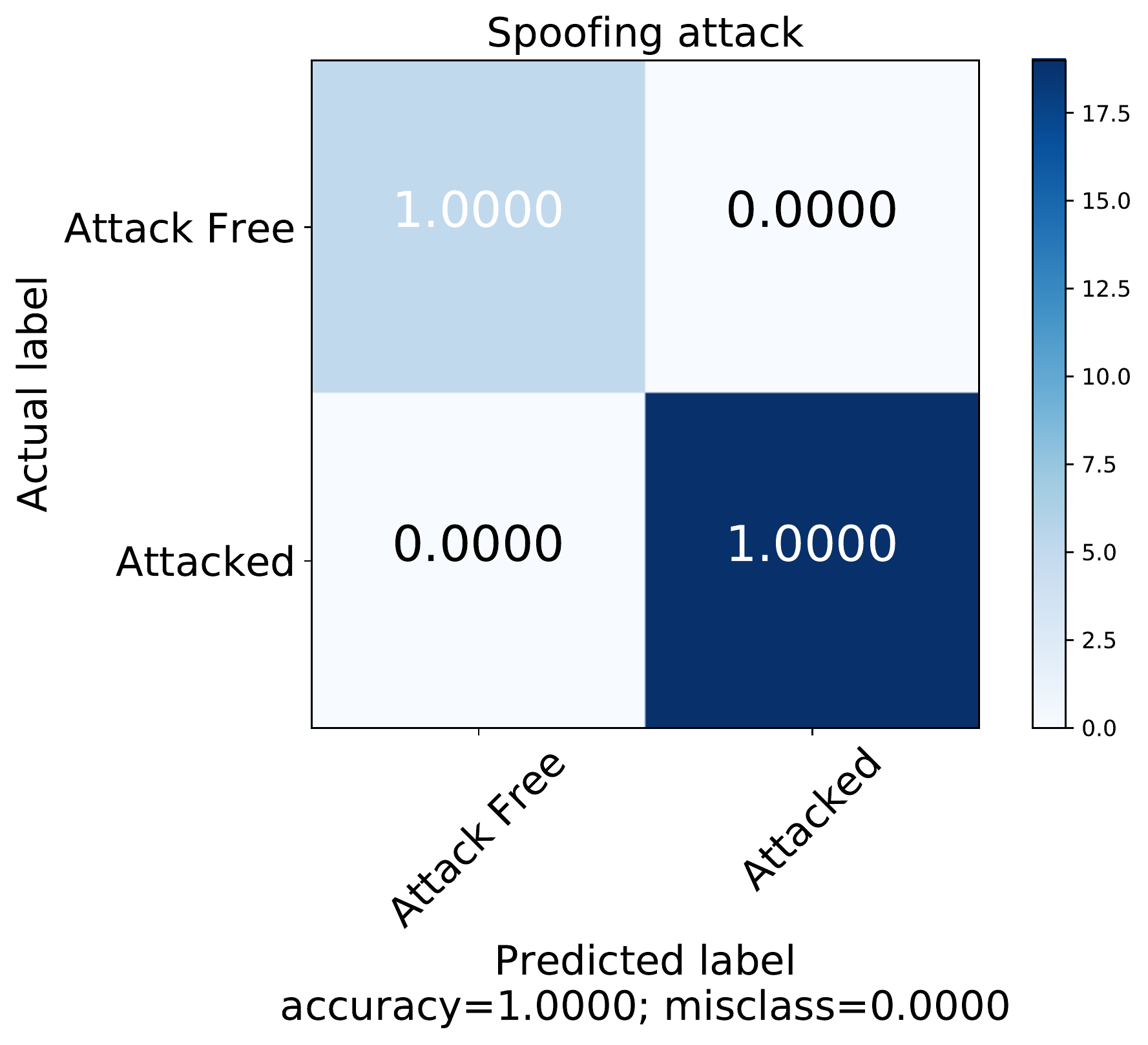}}
    \subfigure[]{\includegraphics[width=0.24\textwidth]{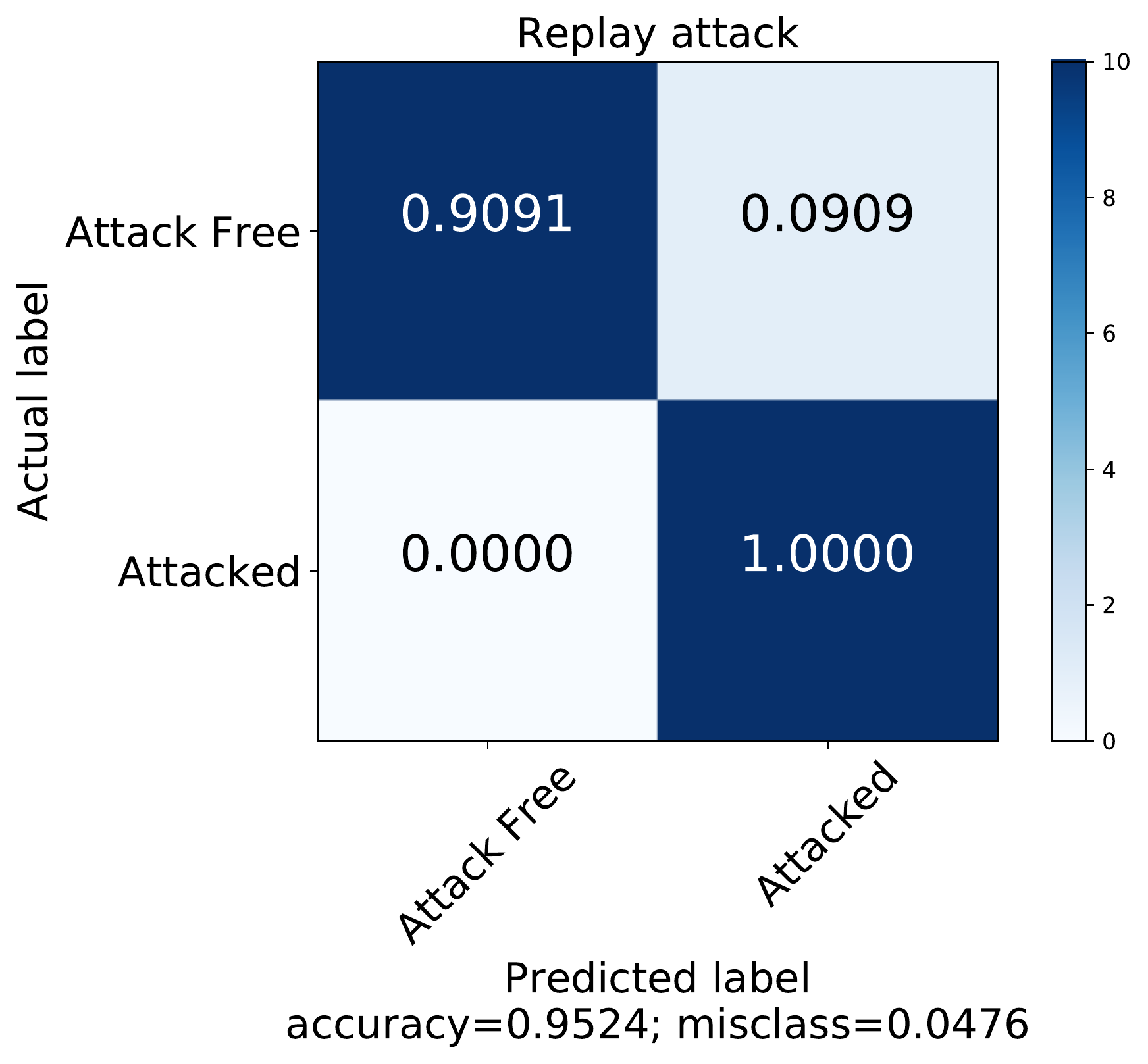}}
    \caption{The proposed methodology has only a 5.26\%, 10\%, 0\%, and 4.76\% misclassification 
rate for DoS, fuzzy, spoofing, and replay attacks, respectively, resulting in only 4.76\% overall misclassification rate.}
    \label{fig:conf_4fig}
    \vspace{-0.5cm}
\vspace{-0.4cm}
\end{figure}

Figure~\ref{fig:conf_4fig} shows the confusion matrix of the proposed methodology. 
We clearly achieve excellent accuracy in detecting
all kinds of attacks described in this section when they are individually tested. The
misclassification rate is very low. According to our analysis, the misclassification 
rate is 5.26\%, 10\%, 0\%, and 4.76\%  for DoS, fuzzy, spoofing, and replay attacks, as shown 
in Figure~\ref{fig:conf_4fig}(a), Figure~\ref{fig:conf_4fig}(b), Figure~\ref{fig:conf_4fig}(c),
and Figure~\ref{fig:conf_4fig}(d), respectively. For
fuzzy-, spoofing-, and replay-attacks, we are able to classify all of the test cases successfully. 
Overall, the proposed methodology has only 4.76\%
misclassification rate DoS-, spoofing-, fuzzy-, and replay-attacked CAN data.

\begin{figure}
    \centering
    \vspace{-0.5cm}
    \subfigure[]{\includegraphics[width=0.24\textwidth]{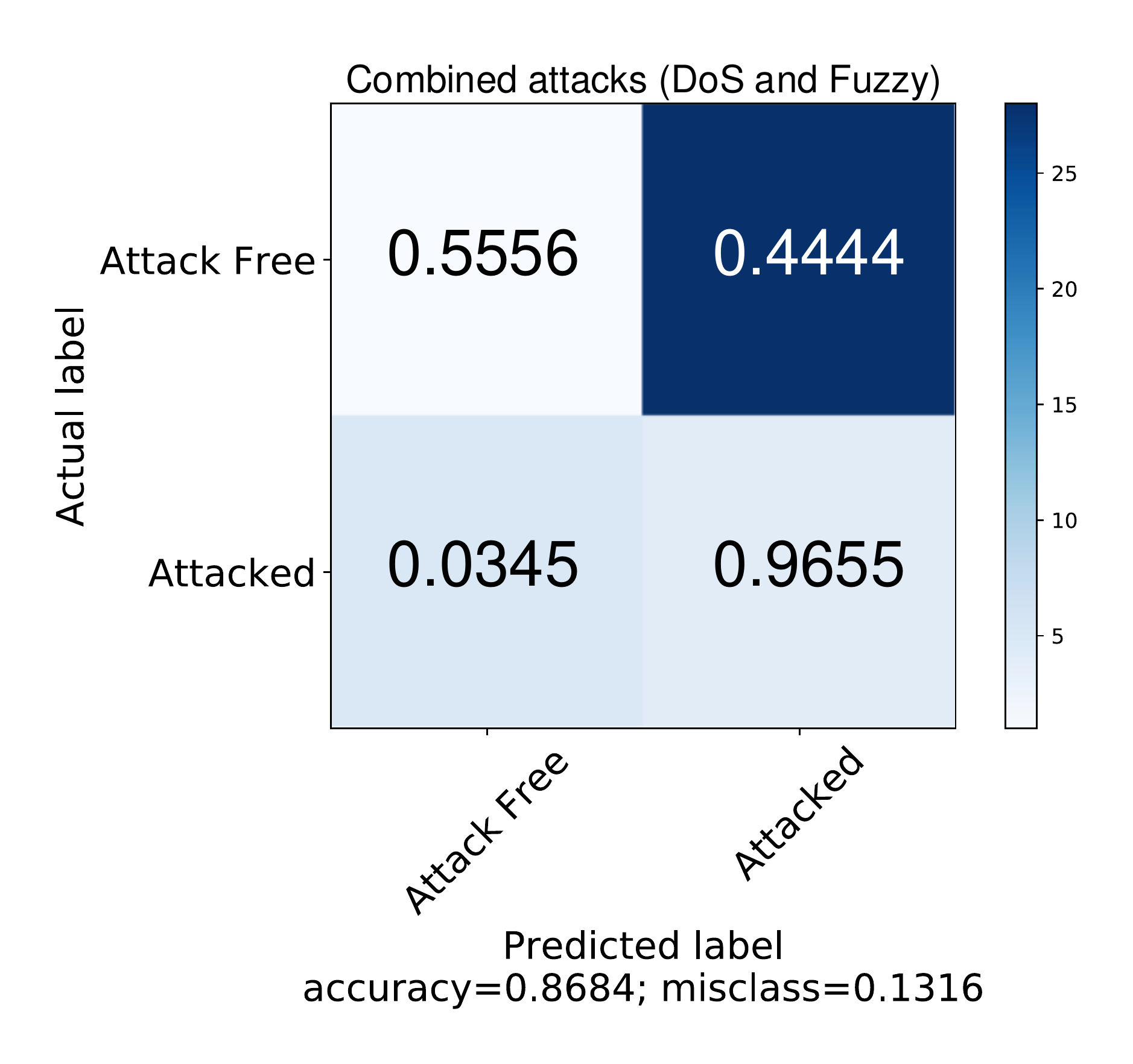}} 
    \subfigure[]{\includegraphics[width=0.24\textwidth]{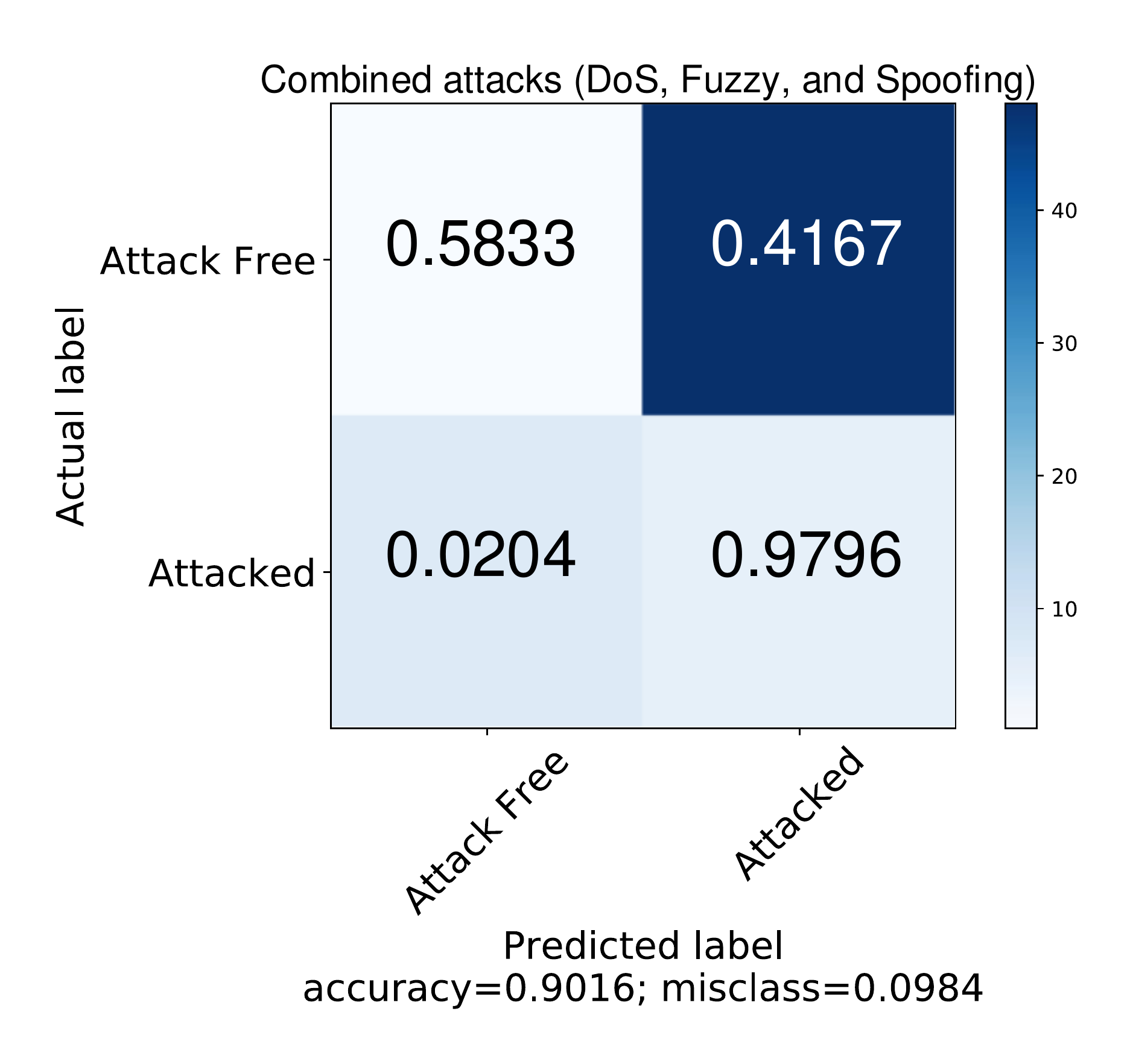}} 
    \caption{The proposed methodology has 13.16\% and 9.84\% misclassification 
rate for combined DoS-fuzzy and DoS-fuzzy-spoofing attacks, respectively.}
    \label{fig:conf_combined_2fig}
    \vspace{-0.5cm}
\end{figure}

We also measure the robustness of the proposed methodology when multiple intruders attack the CAN, simultaneously.
For this analysis, we consider simultaneous DoS and fuzzy attacks and DoS, fuzzy, and spoofing attacks. The misclassification 
rate is 13.16\% for combined DoS and fuzzy attacks, as shown 
in Figure~\ref{fig:conf_combined_2fig}(a). The misclassification 
rate is 9.84\% for combined DoS, fuzzy, and spoofing attacks, as shown 
in Figure~\ref{fig:conf_combined_2fig}(b).

\subsection{Level of Significance Tuning}
\label{subsec:exp_signifi}

The LoS is the probability of
rejecting a hypothesis. It is critical for any decision, since the threshold value changes
depending on the level of significance and $DoF$. We compute 
the $DoF$ using Equation~\ref{eq:chi2}. In our case, we have two 
rows, one for the reference distribution and the other for the 
test distribution. 
The empirical rule states that  
in a normal distribution, 99.7\% of the data points should be 
within (mean $\pm$ $3\times$ (standard deviation)) of the distribution. 
However, researchers have proved that for detecting outliers, considering 
the median in place of the mean gives better results~\cite{Leys:2013}. 
In addition, both the median and the mean signify the central tendency 
of a distribution, but the median does not affected by anomalous data. 
As a result, we considered the median value and divide the reference 
and test distribution into six regions starting from (mean - $3\times$ (standard deviation)) 
to (mean + $3\times$ (standard deviation)) in a step of one standard 
deviation. Hence, our column number is 6. Using the chi-square 
table~\cite{Myatt:2006}, we can chose the LoS given the threshold 
and the degree of freedom.

\begin{figure}
    \centering
    \vspace{-0.6cm}
    \subfigure[]{\includegraphics[width=0.24\textwidth]{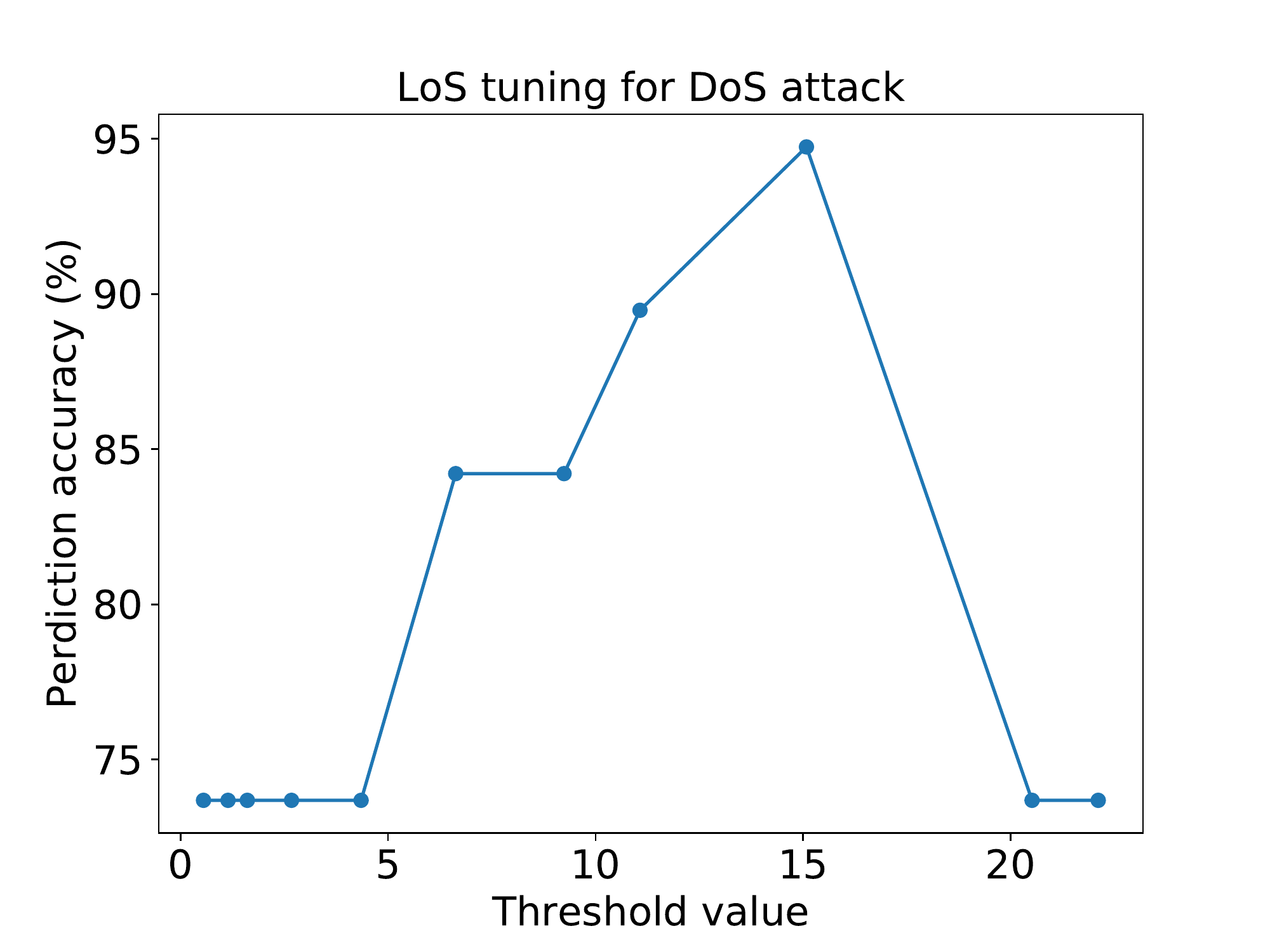}} 
    \subfigure[]{\includegraphics[width=0.24\textwidth]{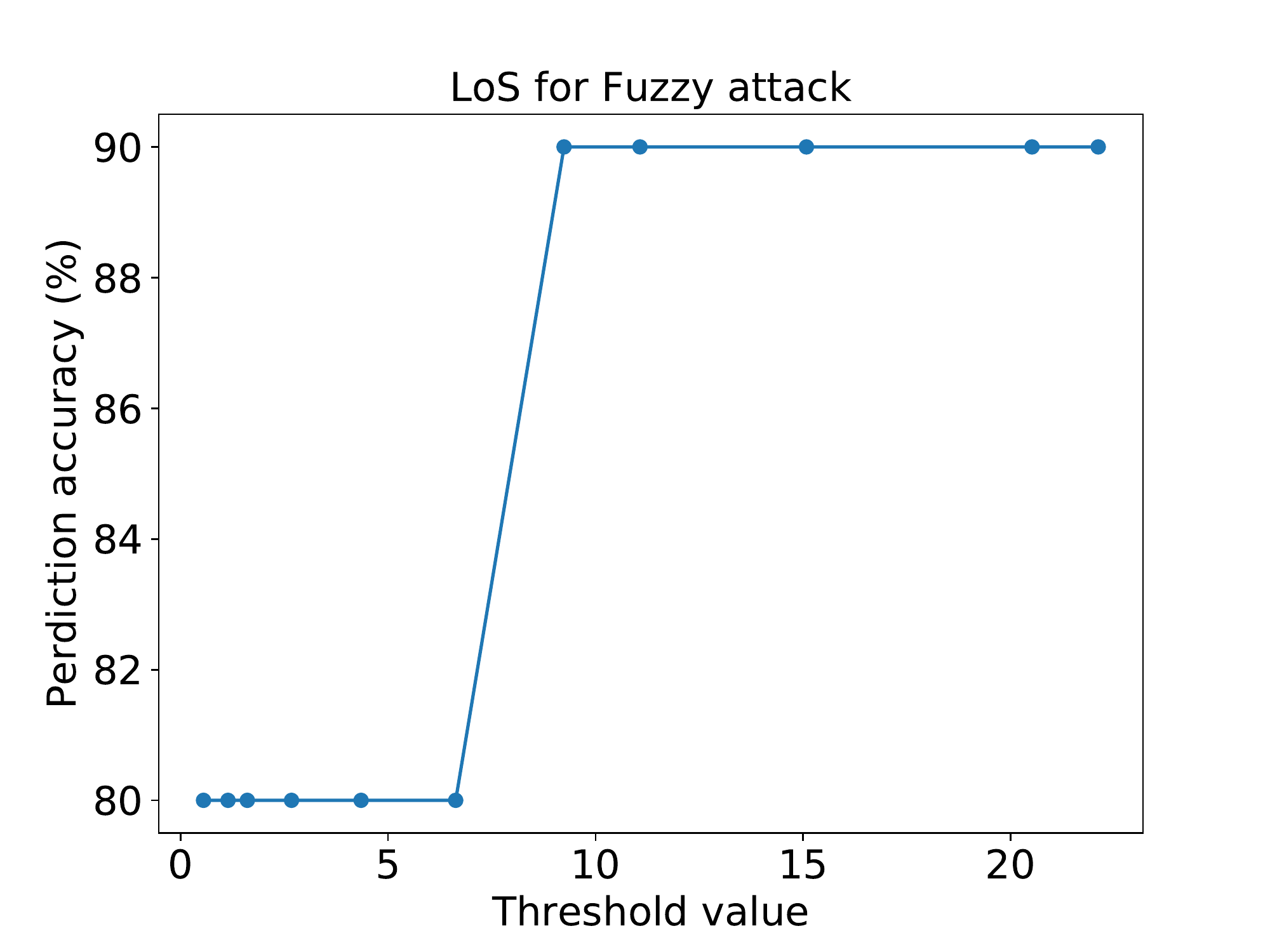}} 
    \subfigure[]{\includegraphics[width=0.24\textwidth]{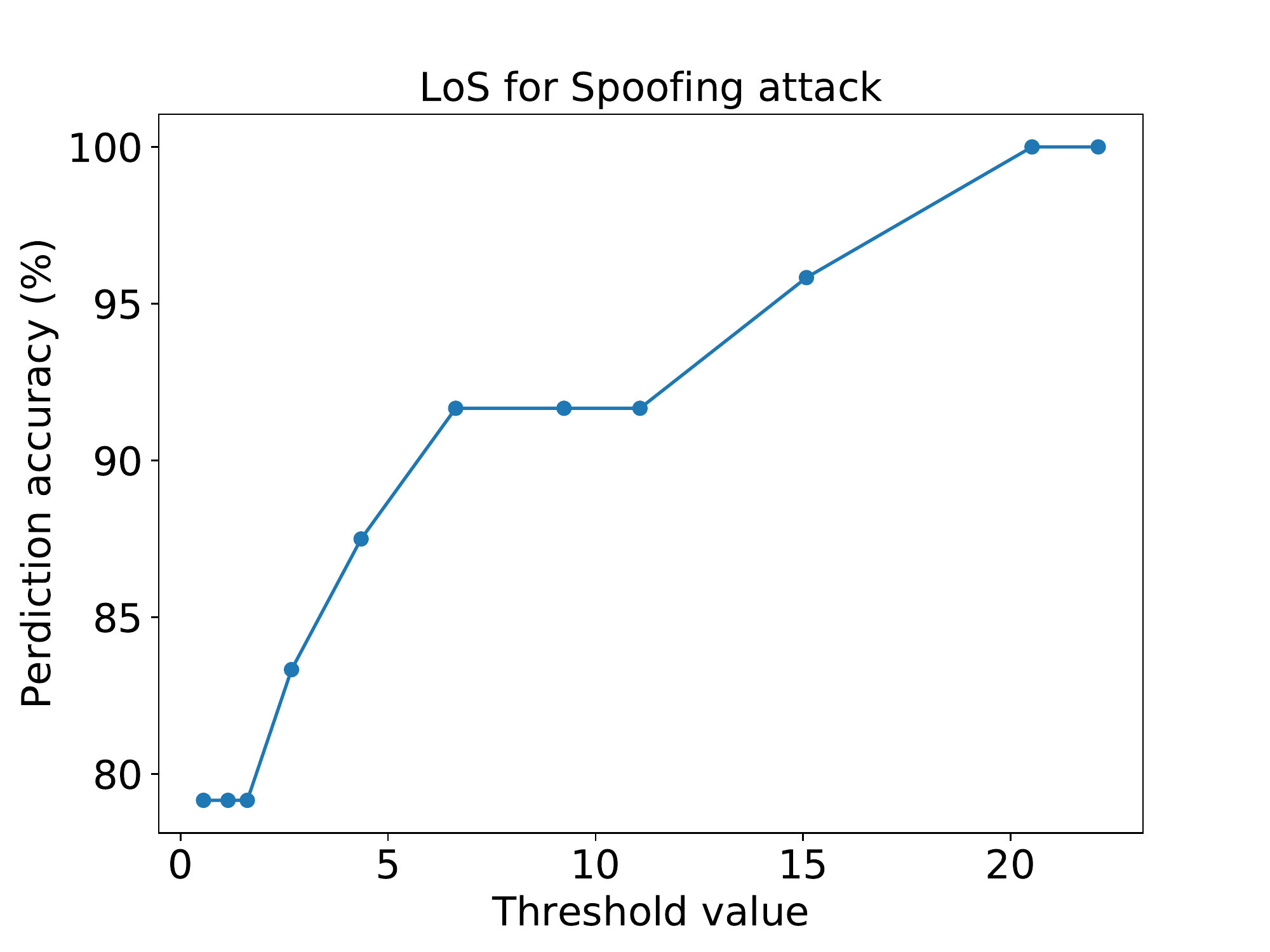}}
    \subfigure[]{\includegraphics[width=0.24\textwidth]{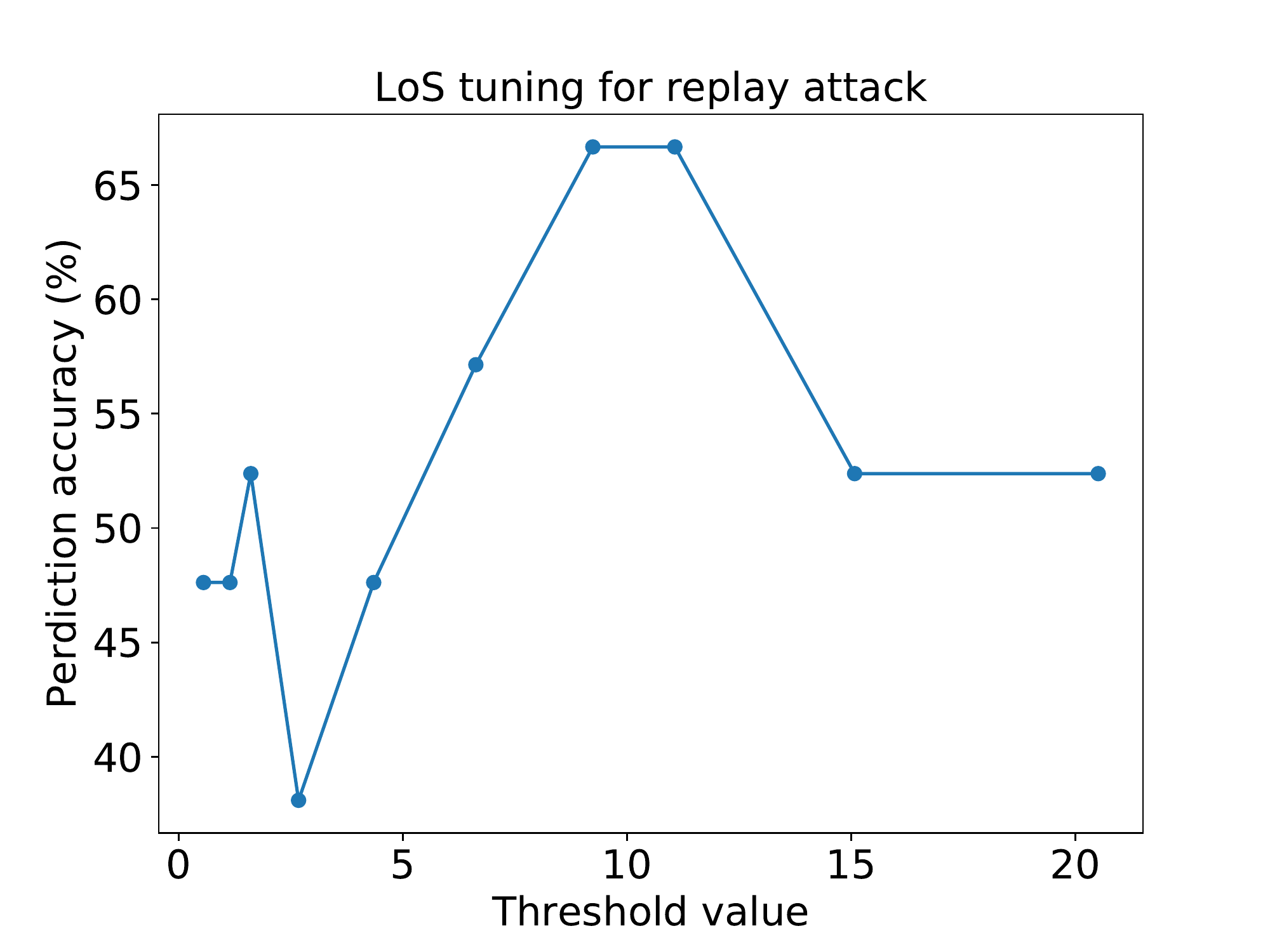}}
    \caption{We identified the best threshold value with level of significance considering the best prediction accuracy.}
    \label{fig:sig_3fig}
\vspace{-0.5cm}
\end{figure}

\begin{figure}
    \centering
    \vspace{-0.3cm}
    \subfigure[]{\includegraphics[width=0.24\textwidth]{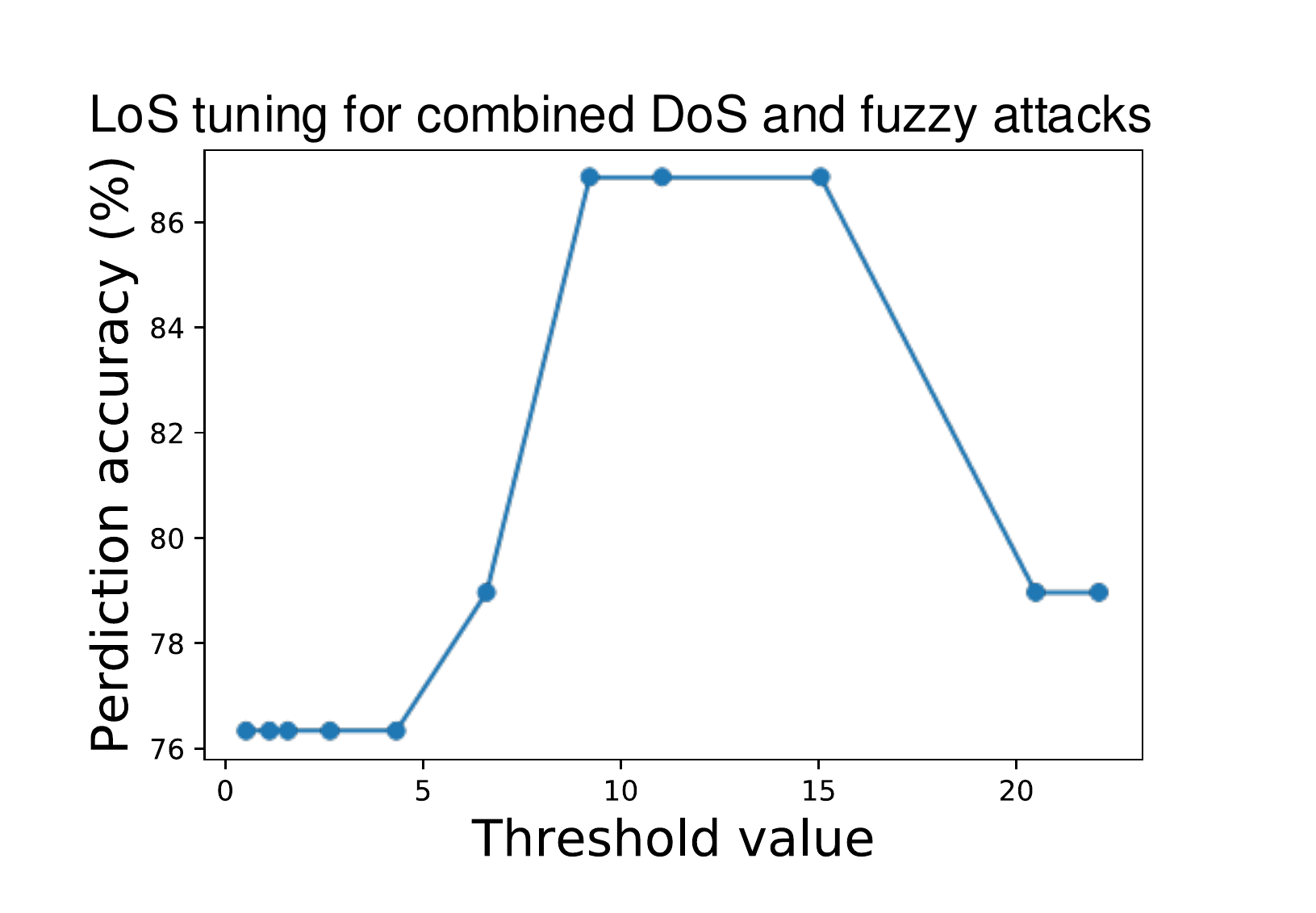}} 
    \subfigure[]{\includegraphics[width=0.24\textwidth]{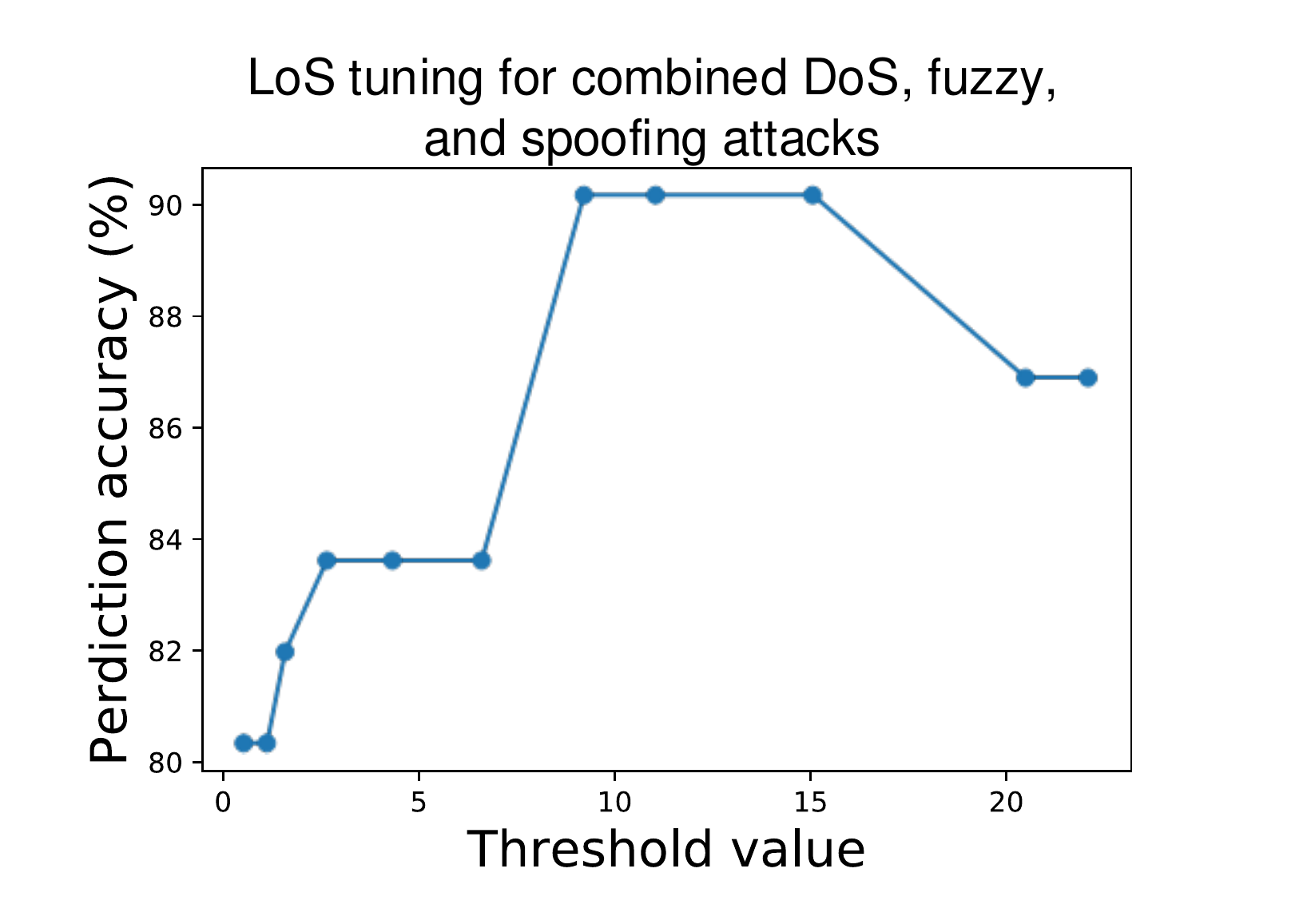}} 
    \caption{Similar to individual attacks, we identified the best threshold value with LoS for combined DoS-fuzzy, and DoS-fuzzy-spoofing attacks considering the best prediction accuracy.}
    \label{fig:sig_com}
\vspace{-0.75cm}
\end{figure}

Our test results show the best LoS we should choose corresponding to the threshold value for comparing
attack-free or attacked distributions. According to our analysis in Section~\ref{subsec:exp_data_ana}, the
distributions of different attacks have different patterns. Hence, we propose a different
level for DoS, fuzzy, spoofing, and replay attacks. Figure~\ref{fig:sig_3fig}(a) suggests 
that a threshold value (15.086) corresponding to a significance level
of 0.01 gives us the best prediction accuracy for a DoS attack. In terms of spoofing and 
fuzzy attacks, we propose a significance level of 0.001 (threshold 20.515) 
and 0.1 (threshold 9.236), as shown in Figure~\ref{fig:sig_3fig}(b) and 
Figure~\ref{fig:sig_3fig}(c), respectively. Using a significance level of 
0.1 and threshold value of 9.236, the proposed chi-squared test method can 
achieve up to 66\% accuracy, as shown in Figure~\ref{fig:sig_3fig}(d). 
However, when we introduce our proposed median test, the prediction 
accuracy increases up to 95.24\% as shown in Figure~\ref{fig:conf_4fig}(d).
Figure~\ref{fig:sig_com}(a) suggests 
that a threshold value (9.236) corresponding to a significance level
of 0.1 gives us the best prediction accuracy of 86.84 for the combined DoS and fuzzy attacks.
Using an LoS of 
0.1 and threshold value of 9.236, the proposed method can 
achieve up to 90.16\% accuracy for combined DoS, fuzzy, and spoofing attacks, as shown in Figure~\ref{fig:sig_com}(b).

\subsection{Comparison to Related Work}
\label{subsec:comparison_related_Work}
We evaluated the efficiency of our approach to detect an attack in CAN's message by 
comparing it to one of the state-of-the-art IDS~\cite{Marchetti:2017}. 
To the best of our knowledge, this is the only methodology that uses CAN bus 
message sequence to identify CAN attacks. Our method is also constructing a 
graph to find a pattern among CAN bus message arbitration IDs and using it 
to detect an attack. Therefore we implemented their approach in the same 
experimental environment using the same real vehicular CAN message dataset~\cite{Lee_data:2018} 
to estimate the effectiveness of our approach. 

\begin{figure}[t!]
	\centering
	\vspace{-0.50cm}
	\includegraphics[width = 0.35\textwidth]{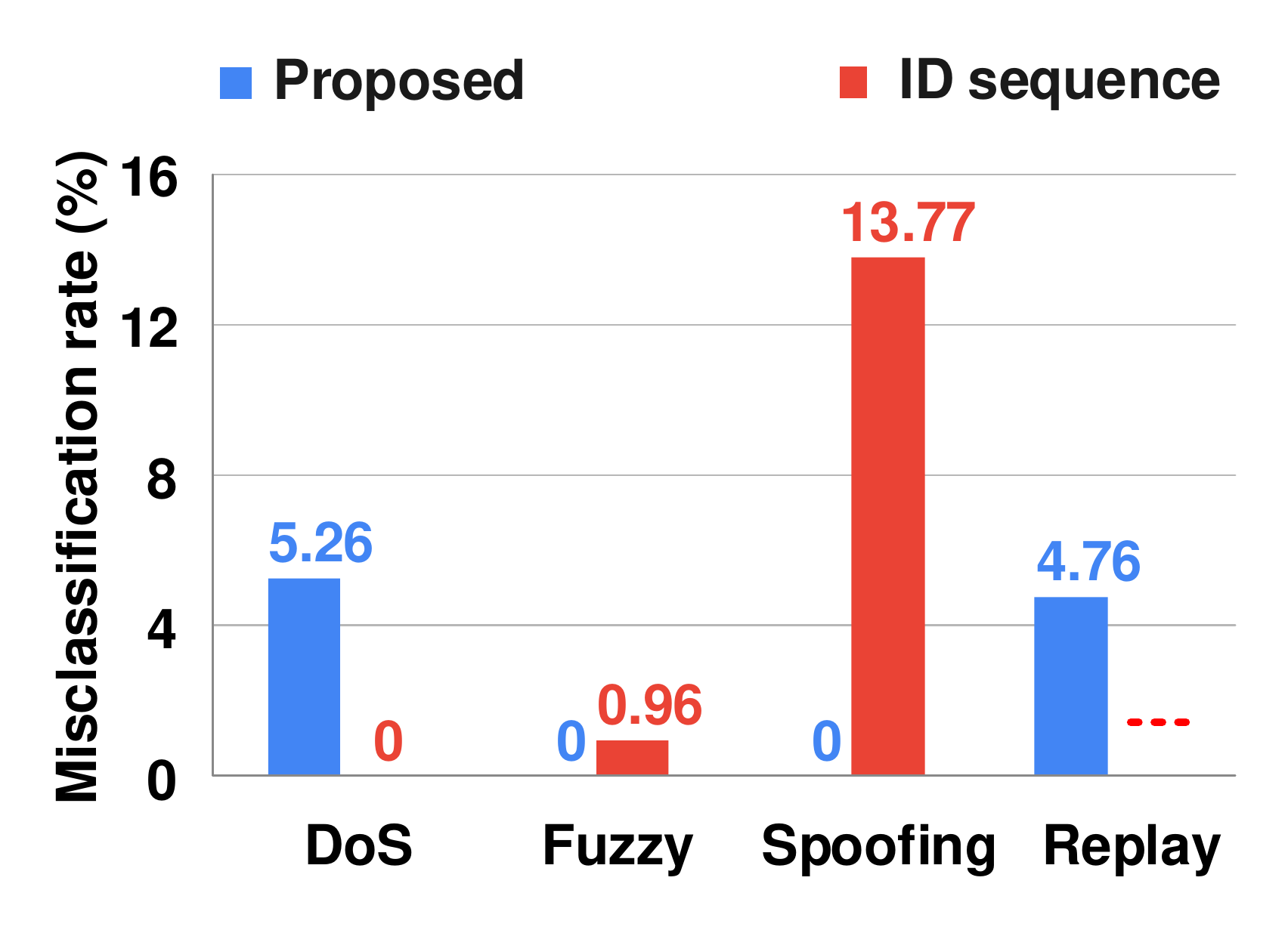}
	\caption {Considering all four-types of attacks, the proposed IDS has a maximum misclassification rate of 5.26\% while the state-of-the-art has 13.77\% and not applicable to replay attacks.}
	\label{fig:mic_class}
	\vspace{-0.50cm}
\end{figure}

In the dataset, the attacker 
targeted the revolutions per minute of an actual vehicle to perform a spoofing 
attack. When we considered spoofing 
attack, the proposed methodology has 13.73\% better accuracy compared to the 
existing method, as shown in Table~\ref{tab:comparison}. When we considered a fuzzy attack, 
the proposed method exhibits comparable accuracy, however, for DoS attack, 
the proposed methodology shows 5.26\% lower accuracy to the existing ID sequence method.
One of the most attractive features of the proposed method is it can detect replay 
attack with 95.24\% accuracy, while the existing method could not detect any replay attacks.
Figure~\ref{fig:mic_class} shows the misclassification rate of 
proposed method compared with the state-of-the-art~\cite{Marchetti:2017}.
The current approach requires much lower computation time 
compared to the proposed method due to the more straightforward implementation. 
However, the proposed methodology requires only up to 258.9$\mu s$ to detect an attack.

\begin{table} [t!] \large 
	\renewcommand{\arraystretch}{1.5}
	\caption{The proposed methodology can successfully detect all four (i.e., DoS-, fuzzy-, spoofing-, and replay) kinds of attacks with reasonable accuracy; however, existing ID sequence methodology can not detect any replay attacks.}
	\label{tab:comparison}
	\centering
	\scalebox{0.65}{
		{\begin{tabular}{|c|c|c|c|c|c|c|c|}
				\hline
				\multirow{2}{*}{Type} & \multirow{2}{*}{Method}           & \multirow{2}{*}{Accuracy}   &  \multirow{2}{*}{TPR}    & \multirow{2}{*}{FPR }     & \multirow{2}{*}{TNR} & \multirow{2}{*}{ FNR } & \multirow{2}{*}{Time}\\
				 & &  (\%) &  (\%) & (\%) & (\%) & (\%) & ($\mu s$)\\
				\hline
				\multirow{4}{*}{ \rotatebox[origin=c]{90}{DoS} }& \multirow{2}{*}{ID sequence}           & \multirow{2}{*}{100}   &  \multirow{2}{*}{99.12}    & \multirow{2}{*}{100 }     & \multirow{2}{*}{0.88} & \multirow{2}{*}{ 0 } & \multirow{2}{*}{ 4.2 }\tabularnewline
				& (\cite{Marchetti:2017}) &  &  &   & &  &  \tabularnewline
				\cline{2-8}
				& \multirow{2}{*}{Proposed}           & \multirow{2}{*}{94.74}   &  \multirow{2}{*}{100}    & \multirow{2}{*}{92.86}     & \multirow{2}{*}{0 } & \multirow{2}{*}{ 7.14 } & \multirow{2}{*}{ 217.5} \\
				& (LoS = 0.01) &  &   &   &  &  &  \\
				\hline
				\hline
                                \multirow{4}{*}{  \rotatebox[origin=c]{90}{Fuzzy} }& \multirow{2}{*}{ID sequence}           & \multirow{2}{*}{99.04}   &  \multirow{2}{*}{98.97}    & \multirow{2}{*}{99.39 }     & \multirow{2}{*}{1.03} & \multirow{2}{*}{ 0.61 } & \multirow{2}{*}{ 3.2 }\tabularnewline
				& (\cite{Marchetti:2017}) &  &  &   & &  &  \tabularnewline
				\cline{2-8}
				& \multirow{2}{*}{Proposed}           & \multirow{2}{*}{100}   &  \multirow{2}{*}{100}    & \multirow{2}{*}{0}     & \multirow{2}{*}{100 } & \multirow{2}{*}{ 0 } & \multirow{2}{*}{ 165.7} \\
				& (LoS = 0.1) &  &   &   &  &  &  \\
				\hline
                                \hline
				\multirow{4}{*}{  \rotatebox[origin=c]{90}{Spoofing}  }& \multirow{2}{*}{ID sequence}           & \multirow{2}{*}{86.23}   &  \multirow{2}{*}{99.3}    & \multirow{2}{*}{53.93 }     & \multirow{2}{*}{46.07} & \multirow{2}{*}{ 0.7 } & \multirow{2}{*}{ 5 }\tabularnewline
				& (\cite{Marchetti:2017}) &  &  &   & &  &  \tabularnewline
				\cline{2-8}
				& \multirow{2}{*}{Proposed}           & \multirow{2}{*}{100}   &  \multirow{2}{*}{100}    & \multirow{2}{*}{0}     & \multirow{2}{*}{100 } & \multirow{2}{*}{ 0 } & \multirow{2}{*}{ 258.9} \\
				& (LoS = 0.001) &  &   &   &  &  &  \\
				\hline
                                \hline
				\multirow{4}{*}{  \rotatebox[origin=c]{90}{Replay}  }& \multirow{2}{*}{ID sequence}           & \multirow{2}{*}{--}   &  \multirow{2}{*}{--}    & \multirow{2}{*}{-- }     & \multirow{2}{*}{--} & \multirow{2}{*}{ -- } & \multirow{2}{*}{ -- }\tabularnewline
				& (\cite{Marchetti:2017}) &  &  &   & &  &  \tabularnewline
				\cline{2-8}
				& \multirow{2}{*}{Proposed}           & \multirow{2}{*}{95.24}   &  \multirow{2}{*}{90.91}    & \multirow{2}{*}{100}     & \multirow{2}{*}{9.01 } & \multirow{2}{*}{ 0 } & \multirow{2}{*}{ 225.7} \\
				& (LoS = 0.1) &  &   &   &  &  &  \\
				\hline
	\end{tabular}}}
	\vspace{-0.70cm}
\end{table}

%% file: conclusion.tex
\section{Conclusion}
\label{sec:conclusion}

As the involvement of modern technologies in the vehicular industry is increasing the 
number of cyber threats, we very much need a robust security mechanism to detect them. 
In this paper, we analyzed the characteristics of all kinds of CAN monitoring-based attacks and 
proposed a four-stage IDS with the help of graph theory, 
statistical analysis, and the chi-square method. 

To the best of our knowledge, this is 
the first graph-based IDS for CAN bus communication.
Our experimental results show that we have a very low misclassification 
rate in detecting attacks or attack free data. In terms of DoS, it is only 5.26\%; for 
the fuzzy attack, it is 10\%; for replay attack, it is 4.76\%; 
and finally, for a spoofing attack, we were able to detect all the malicious attacks. 
The proposed methodology exhibits up to 13.73\% better accuracy compared to 
existing ID sequence-based methods~\cite{Marchetti:2017}.
For strong 
replay attacks, we clearly were not only able to find an attack when it occurred for an 
infected CAN arbitration ID, but also were able to mark the uninfected arbitration IDs as safe. 
In this work, we considered only the distribution of the edge and 
the maximum degree of occurrence of the graph to identify attacks.

In the future, we would like to consider other graph properties such as distribution of 
nodes, cycles, weighted edges etc. In addition, we will apply different machine 
learning algorithms in place of the chi-square test to identify anomalies.